\documentclass[%
 reprint,
superscriptaddress,
 amsmath,amssymb,
 aps,
nofootinbib]{revtex4-2}

\usepackage{color}
\usepackage{hyperref}
\usepackage{graphicx}% Include figure files
\usepackage{dcolumn}% Align table columns on decimal point
\usepackage{bm}% bold math
\begin{document}

%\preprint{APS/123-QED}

\title{ Testing growth rate dependence in cosmological perturbation theory \\
using scale-free models}% Force line breaks with \\
%\thanks{A footnote to the article title}%
\author{Azrul Pohan}\email{azrul.pohan@lpnhe.in2p3.fr}
\affiliation{Laboratoire de Physique Nucléaire et de Hautes Energies, UPMC IN2P3 CNRS UMR 7585, Sorbonne Université, 4, place Jussieu, 75252 Paris Cedex 05, France}
\affiliation{Laboratory of Theoretical Physics, Institut Teknologi Sumatera, Lampung 35365, Indonesia}
\author{Michael Joyce}%
\affiliation{Laboratoire de Physique Nucléaire et de Hautes Energies, UPMC IN2P3 CNRS UMR 7585, Sorbonne Université, 4, place Jussieu, 75252 Paris Cedex 05, France}
%\email{joyce@lpnhe.in2p3.fr}
\author{David Benhaiem}
\affiliation{Laboratoire de Physique Nucléaire et de Hautes Energies, UPMC IN2P3 CNRS UMR 7585, Sorbonne Université, 4, place Jussieu, 75252 Paris Cedex 05, France}
\author{Francesco Sylos Labini}
\affiliation{Centro Ricerche Enrico Fermi,  Via Panisperna 89a, I-00184, Roma, Italy}

%\date{\today}% It is always \today, today,
             %  but any date may be explicitly specified

\begin{abstract}
We generalize previously derived analytic results for the one-loop power spectrum (PS) in  
scale-free models (with linear PS $P(k) \propto k^n$) to a broader class of such models in which 
part of the matterlike component driving the Einstein de Sitter expansion does not cluster. These
models can be conveniently parametrized by $\alpha$, the constant logarithmic 
linear growth rate of fluctuations (with $\alpha=1$ in the usual case).
For $-3< n<-1$, where the one-loop PS is both infrared and ultraviolet convergent and thus
explicitly self-similar, it is characterized conveniently by a 
single numerical coefficient $c(n, \alpha)$.  We compare the analytical predictions for 
 $c(n=-2, \alpha)$ with results from a suite of $N$-body simulations with $\alpha \in [0.25, 1]$ 
performed with an appropriately modified version of the {\sc GADGET} code. Although the 
simulations are of small ($256^3$) boxes, the constraint of self-similarity allows 
the identification of the converged PS at a level of accuracy sufficient to test 
the analytical predictions for the $\alpha$ dependence of the evolved PS.  
Good agreement for the predicted dependence on $\alpha$ of the PS 
is found.  To treat the UV sensitivity of results which grows
as one approaches $n =-1$, we derive exact results incorporating a regularization 
$k_c$ and obtain expressions for $c(n, \alpha, k_c/k)$. Assuming that this 
regularization is compatible with self-similarity allows us to infer a predicted 
functional form of the PS equivalent to that derived in effective field theory (EFT). 
The coefficient of the leading EFT correction at one loop has a strong dependence 
on $\alpha$, with a change in sign at $\alpha \approx 0.16$, providing a  
potentially stringent test of EFT.
\end{abstract}
\maketitle

\section{Introduction}
\label{sec:intro}
Cosmological perturbation theory (PT) is a very important tool in the theory of cosmological structure formation (for a review, see e.g. \cite{Bernardeau2002}).
%\cite{Juszkiewicz1981, Vishniac1983, Goroff1986, Suto1991,makino1992, %Bertschinger1994,bernardeau1993nonlinear,bernardeau1993skewness,catelan1995eulerian,Bernardeau2002}). 
It is essentially the only useful analytical instrument currently available to provide insight into nonlinear dynamics, 
and also an exact benchmark for numerical simulations. Despite its apparent simplicity, it has remained an active area of research over several decades, and there are still open unresolved issues relevant to its application to standard cosmological models. In particular much research has been focused on the sensitivity of the functions describing nonlinear corrections at a given (weakly nonlinear) scale  to contributions from smaller scales. These  ``ultraviolet''
contributions are associated with apparently unphysical divergences in the simplest formulation of PT, and a number of different approaches have been proposed to regulate them  (see e.g. \cite{Baumann2012Cosmo,carrasco2012effective,taruya2012direct,pajer2013renormalization,
blas2014cosmological,hertzberg2014effective,mercolli2014velocity,porto2014lagrangian,carroll2014consistent,carrasco2014effective,carrasco20142,senatore2014redshift,senatore2015ir,baldauf2015bispectrum,vlah2015lagrangian,Angulo2015OneLoop,Baldauf2015EFTTwoLoop,crocce2006renormalize,foreman2016eft,steele2021precise,Wang2022PT,Garny2023PT}). 

Scale-free models, on the other hand, are a family of simplified cosmological models with initial fluctuations characterized by a power spectrum (PS) and an Einstein de Sitter (EdS) expansion law $a(t) \propto t^{2/3}$. Scale-free models are of interest in the context of perturbation theory   ---and more generally ---because they provide a very well-controlled framework within which to understand and test it against numerical results. This is the case because of the so-called {\it self-similar evolution} characterizing these models, which makes the temporal evolution of clustering statistics essentially trivial as it is given by a rescaling of the spatial 
coordinates. This property means that any theoretical predictions which can be made for them will take a much simpler form than in a realistic [e.g. Lambda cold dark matter (LCDM)] cosmology. In perturbation theory, for example, the correction to the PS at each order in perturbation is given by a single number, rather than by a function of scale as in standard models. Further, as has been demonstrated recently \cite{joyce2021quantifying, maleubre2022accuracy}, this same property of self-similarity allows one to obtain very precise results for statistics from numerical simulations. These models can thus provide a potential test-bed for PT and in particular for the question of their ultraviolet divergences and their regulation. 

Scale-free models are usually understood to correspond to a standard EdS cosmology, with source for the expansion being the matter which clusters start from initial Gaussian fluctuations with a PS $P(k) \propto k^n$. This means that one can explore the properties of clustering ---and the adequacy of perturbation theory in describing them ---as a function of the initial conditions (i.e. of $n$), but only within the setting of the single EdS cosmology. In this article we consider perturbation theory in a broader class of scale-free models first considered in \cite{benhaiem2014self} and which we call here {\it generalized scale-free models}. In these models the initial fluctuations are still defined by a power-law PS $P(k) \propto k^n$ but the EdS expansion is driven by the energy density of the clustering matter and, additionally, of a smooth matterlike component (with energy density scaling as $1/a^3$). The EdS model is thus one of a one-parameter family of such models. This parameter can be given by the ratio of the energy density of the matter clustering matter to the 
total energy density,  or equivalently, by the  linear growth rate of density fluctuations. This allows us to potentially exploit the nice properties of scale-free models to test perturbation theory in a broader setting which probes also dependence on the  expansion history, and specifically on the linear growth rate of fluctuations.  We focus here on the simplest canonical analysis in perturbation theory, of the one-loop PS. Building on our derivation in \cite{Michael2022PT} (hereafter P1)  of the kernels in Eulerian and Lagrangian perturbation theory for the generalized EdS cosmologies, we generalize existing analytical results in standard perturbation theory for the one loop PS in the usual scale-free models to these generalized scale-free models.
We analyse the interesting and nontrivial predicted dependences on the growth rate and report some tests of these results against analysis of data from $N$-body simulations performed with an appropriately modified code developed in \cite{benhaiem2014self}. We also discuss how the effective field theory (EFT) approach to the regularization of ultraviolet divergences is modified in this class of scale-free models and the interesting possible numerical tests these results suggest.

\section{Power spectrum in generalized scale-free models}\label{PS-in-GSF}

We consider (as in \cite{benhaiem2014self}) models of pressureless matter clustering under its self-gravity starting from 
density fluctuations which are Gaussian and characterized by a power-law PS $P(k) \propto k^n$.  The expanding cosmological background in 
which it evolves is given by 
\begin{equation}
 H^2= \kappa^2 \frac{8\pi G}{3} \rho_m  
\end{equation}
where $\rho_m$ is the density of clustering matter, $H$ is the Hubble expansion rate, and $\kappa^2 $ is a positive constant.   While the physical interpretation of this expansion law is not in practice of any relevance to our considerations here,  we note that, as discussed in P1 (see also \cite{benhaiem2014self}), for $\kappa^2>1$ one can interpret it 
as arising from the contribution of an additional matter component that does not cluster, while for any $\kappa^2$ it can 
be interpreted in terms of a change in the effective Newton constant governing expansion relative to that governing clustering. 
Doing the standard analysis of linear perturbation theory using this expansion law we obtain a growth law $D(a) \propto a^\alpha$
where the constant growth rate $\alpha$ is related to $\kappa^2$ by the relation
\begin{equation}
\alpha=-\frac{1}{4}+\frac{1}{4}\sqrt{1+\frac{24} {\kappa^2}}\,.\label{eq: alpha}
\end{equation}
Just as in the usual EdS model (with $\kappa^2=1$ and $\alpha=1$), we have an expansion law $a(t) \propto t^{2/3}$ and there is only one characteristic length scale associated with the power-law PS. The property of self-similarity of evolution of 
clustering follows if such evolution is indeed well-defined without cutoffs in the infrared and ultraviolet. Theoretical 
analysis (see e.g. \cite{Peebles1980LSS}) suggest that this can be expected to be true for $-3<n<4$, and many
different studies using numerical simulations indicate that such self-similarity is indeed observed in at least
up to $n=2$ (see e.g. \cite{efstathiou1988gravitational,padmanabhan1995pattern,Colombi1996Self,jain1996self,jain1998self,smith2003stable, Orban2011Self,benhaiem2014self}), and irrespective of whether cosmological EdS expansion is supposed or not
\cite{Baertschiger2007aGravitational,Baertschiger2007bGravitational,Baertschiger2008Gravitational}. 
Indeed a hypothesis underlying numerical simulation in cosmology is 
that clustering is insensitive to the infrared or ultraviolet cutoffs necessarily introduced by such
method (box size, particle density, force smoothing, etc.).

\subsection{Power spectrum in generalized EdS cosmology}
We define canonically (and as in P1) the PS $P(\vec{k}) \equiv P(k)$ ($k=|\textbf{k}|$) of the (assumed) 
statistically homogeneous and isotropic stochastic density field by  
\begin{equation}\label{definition-PS}
    \langle \delta(\textbf{k},a)\delta(\textbf{k}',a) \rangle=(2\pi)^3 \delta^{(D)} (\textbf{k}+\textbf{k}')P(k,a),
\end{equation}
%Explicit calculations of one-loop PT for PS. Give table of results as in Bernardeau et al. 
where $\langle \cdots \rangle$ denotes the ensemble average.
We have shown in P1 that, just as for the usual EdS model, the equations describing 
the clustering of matter in the fluid limit, with irrotational velocity, 
can be solved, in generalized EdS models (gEdS), with a separable ansatz for the 
density field:
\begin{equation}
\label{separable_ansatz}
        \delta(\textbf{k},a)=\sum_{i=1}^{\infty}D^{i}(a)\,\delta^{(i)}(\textbf{k}),
\end{equation}
and likewise for the velocity perturbations.
Assuming that the fluctuations are Gaussian at linear order, 
one obtains the PS at one loop  
as 
\begin{equation}\label{power_spectrum}
\textit{P}_{1-\text{loop}}(k,a)=P_{L}(k,a)+2P_{13}(k,a)+P_{22}(k,a),
\end{equation}
where $P_{L}(k,a)$ is the linear power spectrum and the one-loop contributions are
\begin{widetext}
  \begin{eqnarray}
     P_{13}(k,a)&=&3P_{\text{L}}(k,a) \int \frac{d^{3}q}{(2\pi)^3} P_{L}(q,a) F_{3}^{(s)}(\textbf{k},\textbf{q},-\textbf{q}),\label{P13-one-loop-contributions-general}\\
     P_{22}(k,a)&=&2\int \frac{d^{3}q}{(2\pi)^3} P_{\text{L}}(q,a) P_{\text{L}}(|\textbf{k}-\textbf{q}|,a)|F_{2}^{(s)}(\textbf{k}-\textbf{q},\textbf{q})|^{2}\label{P22-one-loop-contributions-general},
 \end{eqnarray}  
where the superscript ``s" indicates that the kernels $F_2$ and $F_3$ are symmetrized 
with respect to their arguments. These expressions are identical to those in the
standard EdS model and the only difference in the gEdS models come through the
modification to the kernels, which (see P1 for detail) are now functions of the 
parameter $\alpha$:   
 \begin{eqnarray}
      \textit{F}_{2}(\textit{\textbf{q}}_{1},\textit{\textbf{q}}_{2})&=&\Big(\frac{1+4\alpha}{1+6\alpha}\Big) \Tilde{\alpha}(\textit{\textbf{q}}_{1},\textit{\textbf{q}}_{2})+\Big(\frac{2\alpha}{1+6\alpha}\Big) \Tilde{\beta}(\textit{\textbf{q}}_{1},\textit{\textbf{q}}_{2}),\\
      \textit{G}_{2}(\textit{\textbf{q}}_{1},\textit{\textbf{q}}_{2})&=&\Big(\frac{1+2\alpha}{1+6\alpha}\Big) \Tilde{\alpha}(\textit{\textbf{q}}_{1},\textit{\textbf{q}}_{2})+\Big(\frac{4\alpha}{1+6\alpha}\Big) \Tilde{\beta}(\textit{\textbf{q}}_{1},\textit{\textbf{q}}_{2}),\\
       \textit{F}_{3} (\textit{\textbf{q}}_{1},\textit{\textbf{q}}_{2},\textit{\textbf{q}}_{3}) &=&\frac{1}{2}\Bigg\{\Big(\frac{1+6\alpha}{1+8\alpha}\Big)\Tilde{\alpha}(\textit{\textbf{q}}_{1},\textit{\textbf{q}}_{2}+\textit{\textbf{q}}_{3})\textit{F}_{2}(\textit{\textbf{q}}_{2},\textit{\textbf{q}}_{3})+\Big(\frac{2\alpha}{1+8\alpha}\Big)\Tilde{\beta}(\textit{\textbf{q}}_{1},\textit{\textbf{q}}_{2}+\textit{\textbf{q}}_{3})\textit{G}_{2}(\textit{\textbf{q}}_{2},\textit{\textbf{q}}_{3})\nonumber\\
       &&+\Big[\Big(\frac{1+6\alpha}{1+8\alpha}\Big)\Tilde{\alpha}(\textit{\textbf{q}}_{1}+\textit{\textbf{q}}_{2},\textit{\textbf{q}}_{3})
      +\Big(\frac{2\alpha}{1+8\alpha}\Big)\Tilde{\beta}(\textit{\textbf{q}}_{1}+\textit{\textbf{q}}_{2},\textit{\textbf{q}}_{3})\Big]
       \textit{G}_{2}(\textit{\textbf{q}}_{1},\textit{\textbf{q}}_{2}) \Bigg\},
 \end{eqnarray}
 where 
 \begin{equation}
   \Tilde{\alpha}(\textit{\textbf{q}}_{1},\textit{\textbf{q}}_{2})=\frac{\textit{\textbf{q}}_{1}.(\textit{\textbf{q}}_{1}+\textit{\textbf{q}}_{2})}{\textit{q}_{1}^{2}}\,, \qquad
   \Tilde{\beta}(\textit{\textbf{q}}_{1},\textit{\textbf{q}}_{2})=\frac{1}{2}(\textit{\textbf{q}}_{1}+\textit{\textbf{q}}_{2})^{2}\frac{\textit{\textbf{q}}_{1}.\textit{\textbf{q}}_{2}}{\textit{q}_{1}^{2}\textit{q}_{2}^{2}}.
  \end{equation} 
\end{widetext}
Using these expressions (see P1) the PS at one loop is then expressed in terms of three 
integrals with $\alpha$ dependent coefficients:
\begin{eqnarray}\label{P22-term}
P_{22}&=&M_0+\frac{1+4\alpha}{1+6\alpha} M_1 + \left(\frac{1+4\alpha}{1+6\alpha}\right)^2 M_2 ,\nonumber \\
2 P_{13}&=&N_0
+ \frac{1+2\alpha}{1+8\alpha} N_1 +\frac{2\alpha(1+2\alpha)}{(1+6\alpha)(1+8\alpha)} N_2\label{P13-term},
\end{eqnarray}
where the $M_i(k)$ are the integrals
\begin{eqnarray}
M_i&=&\frac{1}{8\pi^2}k^{3}\int_{0}^{\infty} d r\int_{-1}^{1} d \mu P_{L} (k r)\nonumber\\
& &\times\frac{P_{L} (k\sqrt{1+r^2 -2\mu r})}{(1+r^2 -2\mu r)^2} m_i(r,\mu),
\end{eqnarray}
with
\begin{eqnarray}
m_0 (r,\mu)&=&(\mu-r)^2,\\%\frac{(\mu-r)^2}{(1+r^2 -2\mu r)^2} \\
m_1 (r,\mu)&=& 4r(\mu-r)(1-\mu^2), \\
m_2 (r,\mu)&=&4r^2 (1-\mu^2)^2,
\end{eqnarray}
and the $N_i(k)$ integrals are 
\begin{equation}
N_i=\frac{1}{8\pi^2}k^{3} P_{L} (k) \int_{0}^{\infty} dr P_{L} (kr) n_i(r),
\end{equation}
with 
\begin{eqnarray}
         n_0&=&-\frac{4}{3}, \\
         n_1&=&1+\frac{8}{3}r^2-r^4+\frac{(r^{2}-1)^{3}}{2r} \ln\frac{|1+r|}{|1-r|}, \\
         n_2&=&\frac{1}{r^2}\Big(1-\frac{8}{3}r^2-r^4\Big)+\frac{(r^{2}-1)^{3}}{2r^3} \ln\frac{|1+r|}{|1-r|}.
\end{eqnarray}
The variables $r$ and $\mu$ in the integrals have been defined from the momenta in
Eqs.~\eqref{P13-one-loop-contributions-general}-\eqref{P22-one-loop-contributions-general}
as $r=q/k$ and  $\mu=\textbf{k}.\textbf{q}/(kq)$. 

\subsection{Power spectrum for scale-free initial conditions}

We now consider the case that $P_{L}(k)$ is a simple power law.
In order to control carefully for infrared and ultraviolet divergences 
we introduce cutoffs, taking
\begin{equation}
    P_{L}(k,a)=
    \begin{cases}
      A D^{2} k^{n}, & \text{if}\quad\varepsilon \leq k\leq k_{c}  \\
      0, & \text{otherwise}\label{PowerLaw-Input}
    \end{cases}
  \end{equation}
where A is the amplitude of the power spectrum at $a=1$, $D\equiv a^\alpha$ is the linear growth rate of fluctuations, and $\varepsilon$ ($k_{c}$) are the infrared (ultraviolet) cutoffs.

We will work with the dimensionless power spectrum, defined canonically as 
\begin{equation}
    \Delta^{2}(k)=\frac{ k^{3}P(k)}{2\pi^{2}}\label{dimensionless-PS}.
\end{equation}
The one-loop result in Eq.~\eqref{power_spectrum} can then conveniently 
be rewritten as 
\begin{equation}\label{dimensionlessPS-oneLoop}
\Delta_{1-loop}^{2}(k)=\Delta^{2}_{L}\left[1+c\Big(n,\alpha, \frac{\varepsilon}{k}, \frac{k_c}{k} \Big) \Delta^{2}_{L}\right],
\end{equation}
for $\varepsilon \leq k \leq k_c$,  and where $\Delta_{L}^{2}(k)=\frac{ k^{3}P_{L}(k)}{2\pi^{2}}$.

The dimensionless constant $c$ in Eq.~\eqref{dimensionlessPS-oneLoop} is then given by  
\begin{eqnarray}
     c\Big(n,\alpha, \frac{\varepsilon}{k}, \frac{k_c}{k} \Big)&=& \hat{M}_0+\frac{1+4\alpha}{1+6\alpha} \Big[\hat{M}_1 + \frac{1+4\alpha}{1+6\alpha} \hat{M}_2\Big] 
     \nonumber\\
& &+\hat{N}_0 + \frac{1+2\alpha}{1+8\alpha} \Big[\hat{N}_1+ \frac{2\alpha}{1+6\alpha} \hat{N}_2\Big]\nonumber\\
     %+\frac{2\alpha(1+2\alpha)}{(1+6\alpha)(1+8\alpha)} N_2
     \label{c-one-loop}
 \end{eqnarray}
where the $\hat{M}_i$ and $\hat{N}_i$ are dimensionless integrals:
\begin{eqnarray}
\hat{M}_i&=&\frac{1}{4}\int_{\varepsilon/k}^{k_{c}/k} dr \, r^{n} \int_{\mu_{min}}^{\mu_{max}} d \mu (1+r^2 -2\mu r)^{\frac{n}{2}-2}\, m_i(r,\mu)\nonumber\\\\
\hat{N}_i&=&\frac{1}{4} \int_{\varepsilon/k}^{k_{c}/k} dr \, r^n \, n_i(r)
\label{Mhat-Nhat}
\end{eqnarray}
where $m_i$ and $n_i$ are the same functions defined above, 
and 
\begin{eqnarray}
\mu_{min}(r)&=& \text{\rm Max}\Big\{-1,\frac{1+r^2-(k_c/k)^{2}}{2r}\Big\},\nonumber\\
            \mu_{max}(r)&=&\text{\rm Min}\Big\{1,\frac{1+r^2-(\varepsilon/k)^{2}}{2r}\Big\},
    \end{eqnarray}
 are the angular integration limits.

Defining the characteristic scale $k_{NL}$ by  $\Delta^{2}_{L} (k_{NL})\equiv 1$, 
we have 
\begin{equation}
{k_{NL}}(a) \propto a^{-\frac{2\alpha}{3+n} },
\end{equation}
and, given the assumed power-law form,  
\begin{eqnarray}
     \Delta_{L}^2=\Big(\frac{k}{k_{NL}}\Big)^{(n+3)}.
\end{eqnarray}

If $c$ remains finite when we take the limits $\varepsilon \rightarrow 0$ and $k_c \rightarrow \infty$, $c$ becomes a function of $n$ and $\alpha$ only, with
\begin{eqnarray}\label{self-similar-dimensional-reg}
\Delta_{1-loop}^{2}(k)=\Delta^{2}_{L}\left[1+c\Big(n,\alpha\Big) \Delta^{2}_{L}\right].
\end{eqnarray}
The evolution is then explicitly self-similar in a sense that
\begin{equation}\label{self-similar}
\Delta^{2} (k, a) = \Delta^{2} \Big(\frac{k}{k_{NL}(a)},1\Big)
\end{equation}
i.e. the temporal evolution of clustering corresponds to a rescaling of the spatial coordinates in proportion 
to the sole characteristic scale, the nonlinearity scale $\propto k_{NL}^{-1}$, defined by the power-law PS.    

\subsection{Convergence analysis}
\label{Convergence analysis}

By studying the behavior of the integrals $\hat{M}_i$  and $\hat{N}_i$ in the limit
$\varepsilon/k \rightarrow 0$ and $k_c/k \rightarrow \infty$ we can determine their
infrared and ultraviolet convergence properties. Following standard analysis,
and as discussed also in P1, the two dimensional integrals $\hat{M}_i$ have divergences 
for certain cases in the limit $\varepsilon/k \rightarrow 0$ at $r=0$ and $r=1$. 
As noted e.g. by \cite{makino1992} the contribution of each is in fact identical
because of the symmetry of the integrals (the $r=1$ divergence corresponds 
to $|\bf{q}-\bf{k}|\rightarrow 0$, which is identical to the $r=0$ contribution
after a change in variable). This means that the infrared behavior can 
be determined simply by doubling the $r=0$ contribution, which can 
easily be inferred from a Taylor expansion. 

Explicitly the leading behavior as $r \rightarrow 0$ of the integrands of $\hat{M}_0$ and 
$\hat{N}_0$ is $\sim r^n$, leading to divergence for $n\leq -1$,  but when summed 
(and taking into account the factor of two mentioned above) these leading divergences 
cancel and give a ``safe" leading behavior $\sim r^{n+2}$ i.e. convergence for $n\geq -3$. 
The integrands of the four integrals $\hat{M}_1,\hat{M}_2,\hat{N}_1,\hat{N}_2$, which 
contribute to the PS via an $\alpha$-dependent pre-factor,  all have this 
same safe behavior. As noted in P1 the overall infrared convergence for 
any $n >-3$ thus holds for any $\alpha$, exactly as in the standard EdS model.
This result is expected since such convergence is a consequence of Galilean invariance
\cite{scoccimarro1996loop,peloso2013galilean}, a property that is respected by the generalized 
EdS cosmologies just as in the canonical case.

\begin{table}[tpb]
\caption{Expansion around $(1/r)=0$ of the integrands of 
$\hat{M}_{i}$ and $\hat{N}_{i}$. As in the standard EdS model ($\alpha=1$) these imply
that the one loop PS is divergent for $n>-1$. As discussed in the text, the coefficients
of these divergences depend on $\alpha$ and at a specific value ($\alpha \approx 0.16$)
the leading divergence vanishes and the one-loop result remains ultraviolet convergent
for $n<1/2$. 
}

\centering
\begin{tabular}{|c c |} 
 \hline
  &  expansion of integrand    \\ [0.5ex] %  &  value for $n=-1$
 \hline\hline
 $\hat{M}_{0}$ &  $r^{2n-2} \Big[\frac{1}{2}+\frac{n^2-3 n-2}{12 r^2}+O\big(\frac{1}{r}\big)^4\Big]$ \\ %& $-2$
 $\hat{M}_{1}$ & $r^{2n-2} \Big[ -\frac{4}{3}-\frac{2\left(n^2-3 n-4\right)}{15 r^2}+O\big(\frac{1}{r}\big)^4\Big]$  \\ %& $0$ or $-2$ (if $\alpha=0.16\cdots$)
 $\hat{M}_{2}$ & $r^{2n-2} \Big[ \frac{16}{15}+\frac{8 \left(n^2-3 n-4\right)}{105 r^2}+O\big(\frac{1}{r}\big)^4\Big]$  \\ %& $-2$
  $\hat{N}_{0}$ & $-\frac{1}{3}r^{n}$  \\ 
$\hat{N}_{1}$ & $r^{n} \Big[ \frac{4}{5}-\frac{4}{35 r^2}+O\big(\frac{1}{r}\big)^4\Big]$  \\ %& $-3$
  $\hat{N}_{2}$ &  $r^{n} \Big[ -\frac{4}{3}+\frac{4}{5 r^2}+O\big(\frac{1}{r}\big)^4\Big]$   \\ %&  $-5$
  [1ex] 
 \hline
\end{tabular}
\label{Table-UV}
\end{table}

\begin{figure}[t]
\centering\includegraphics[width=8cm, height=8cm]{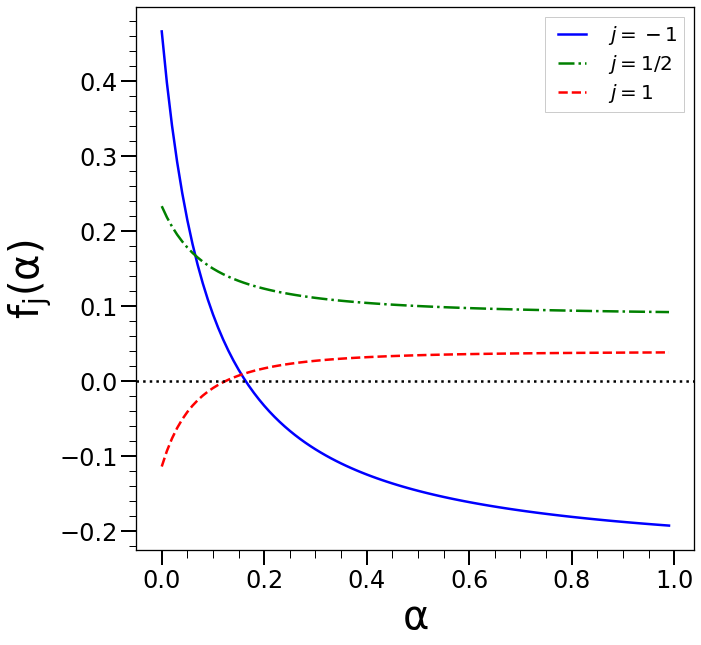}
\caption{The prefactors  in Eqs.~\eqref{c-UVexpansion}  and \eqref{c-UVexpansion-2} for $j=-1,1/2,1$  
as a function of $\alpha$. The associated leading UV contribution, which diverges for $n \geq -1$, 
is proportional to $f_{-1}$ and thus vanishes at $\alpha=\alpha_c \approx 0.16$.}
\label{fj-alpha}
\end{figure}
For $r \rightarrow \infty$, on the other hand, the integrands in $\hat{N}_0$, $\hat{N}_1$, $\hat{N}_2$
all have the same leading behavior $\sim r^{n}$, and all those in $\hat{M}_0$, $\hat{M}_1$, $\hat{M_2}$ 
the leading behavior $\sim r^{2n-2}$. For the canonical $\alpha=1$ case, the one loop PS therefore
diverges for $n>-1$ with a leading divergence coming from the term $\sim r^{n}$ for $n<2$,
and from the term $\sim r^{2n-2}$ for $n>2$. As noted in P1, the same result holds in the gEdS models, 
except for one important difference: the coefficient of the leading divergence vanishes at 
a specific value of $\alpha$. This can be seen by using the results in Table~\ref{Table-UV} to infer the linear combination of these integrands which is used to obtain the
one loop PS as in Eq.~(\ref{c-one-loop}). The expansion 
around $(1/r)=0$ of the resultant integrand is then 
\begin{equation}
    \label{c-UVexpansion}    
     f_{-1}(\alpha)r^n +f_{1/2}(\alpha)r^{2n-2} + O\Big[r^{n-2}, r^{2n-4}\Big]
\end{equation}
with the former giving the leading term for $n<2$ and the latter for $n>2$,
and where
\begin{eqnarray}
\label{UV-factors}    
     f_{-1}(\alpha)&=&\frac{7-14\alpha-176\alpha^2}{15(1+6\alpha)(1+8\alpha)} ,\\ 
    % f_{1}(\alpha)&=&\frac{4(1+2\alpha)(8\alpha-1)}{35(1+6\alpha)(1+8\alpha)} \\ 
    f_{1/2}(\alpha)&=&\frac{7+36\alpha+92\alpha^2}{30(1+6\alpha)^2}\,.\label{UV-factor1/2}
\end{eqnarray}
 The indices of the functions $f$ have been chosen to indicate 
 the value of $n$ at which the corresponding terms lead to ultraviolet 
 divergence of $c$. As noted in P1 the function $f_{-1}$ 
 crosses zero at  $\alpha=\alpha_c$, where
\begin{equation}
\alpha_c=0.1635\cdots   
\end{equation}  
while $f_{1/2}$ is always nonzero and of the same 
sign as in the case $\alpha=1$ (see Fig.~\ref{fj-alpha}). 
Thus the leading divergence actually 
vanishes at this specific value $\alpha_c$, and one loop PT gives 
in this case a well-defined (i.e. finite) prediction up to 
$n=1/2$. The two leading terms in the expansion of
the integrand in $c$ about $(1/r)=0$ are then given by 
\begin{equation}
    \label{c-UVexpansion-2}    
     f_{1}(\alpha_c)r^{n-2} +f_{1/2}(\alpha_c)r^{2n-2} + O\Big[r^{n-4}, r^{2n-4}\Big]
\end{equation}
where 
\begin{equation}
\label{UV-factor-f1}      
    f_{1}(\alpha)=\frac{4(1+2\alpha)(8\alpha-1)}{35(1+6\alpha)(1+8\alpha)}. 
\end{equation}
For $n<0$ the first term is the leading one while for $n>0$ it is the latter.

We will return to discuss these behaviors in more 
detail in Section \ref{UV divergences} below,  in 
which we consider the regularization 
of ultraviolet divergences in these models.
Until then we lay aside the consideration of
these divergences, deriving exact
one loop results for the ultraviolet 
convergent regime (for any $\alpha$ i.e.
for $n<-1$). We report our numerical
tests of these results,  in the still
more restricted regime where they
appear to be very insensitive to 
(finite) contributions from  
ultraviolet scales.

\subsection{Exact results for PS ($-3< n<-1$)}

To obtain an analytical expression for the one-loop corrections in the range where there are the infrared divergences 
cancel out and there are no ultraviolet  divergences, i.e. for $-3<n<-1$, it is convenient to use 
dimensional regularization to treat the infrared divergences in the individual contributing terms (as in \cite{scoccimarro1996loop,pajer2013renormalization}).
To do so, it is convenient to work directly with the initial unsimplified expressions for $P_{13}$ and $P_{22}$ 
as in Eqs.~\eqref{P13-one-loop-contributions-general} and \eqref{P22-one-loop-contributions-general} 
where $P_{L}$ is a simple power-law (without cutoffs).
%refers to Eq.~\eqref{PowerLaw-Input}.
Replacing the integrations $\int d^3 q$ by $\int d^d q$ 
we obtain
\begin{eqnarray}
     P_{22}(k,a)&=&A^{2}a^{4\alpha}\int \frac{d^{d}\textbf{q}}{(2\pi)^3} q^{n}2|\textbf{k}-\textbf{q}|^{n}| F_{2}^{(s)}(\textbf{k}-\textbf{q},\textbf{q})|^{2}\label{Integration-P22-dimenreg},\nonumber\\ \\
     P_{13}(k,a)&=&A^{2}a^{4\alpha}\int \frac{d^{d}\textbf{q}}{(2\pi)^3} 3q^{n}k^{n}F_{3}^{(s)}(\textbf{k},\textbf{q},-\textbf{q})\label{Integration-P13-dimenreg}.
\end{eqnarray}
To integrate Eqs.~\eqref{Integration-P22-dimenreg} and \eqref{Integration-P13-dimenreg} we use the formula (see the appendix in \cite{scoccimarro1996loop}) as below
    \begin{eqnarray}\label{formula-PS-dimenreg}
         & &\int \frac{d^{d}\textbf{q}}{(q^2)^{\nu_1}[(\textbf{k}-\textbf{q})^2]^{\nu_2}}\nonumber\\
         &=&\frac{\Gamma(d/2-\nu_1)\Gamma(d/2-\nu_2)\Gamma(\nu_1+\nu_2-d/2)}{\Gamma(\nu_1)\Gamma(\nu_2)\Gamma(d-\nu_1-\nu_2)}\nonumber\\
         & &\times\pi^{d/2}k^{d-2\nu_1-2\nu_2}
 \end{eqnarray}
 together with relation 
 \begin{equation}
     \textbf{k}\cdot \textbf{q}=\pm\frac{1}{2}(k^{2}+q^{2}-|\textbf{k}\mp\textbf{q}|^2).
 \end{equation}
 \begin{figure*}
 \includegraphics[width=8cm,height=8cm]{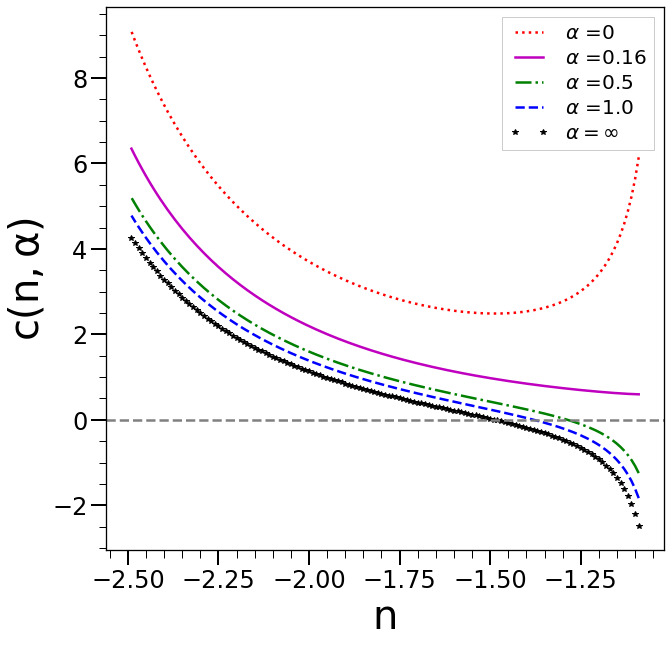}
    \includegraphics[width=8cm,height=8cm]{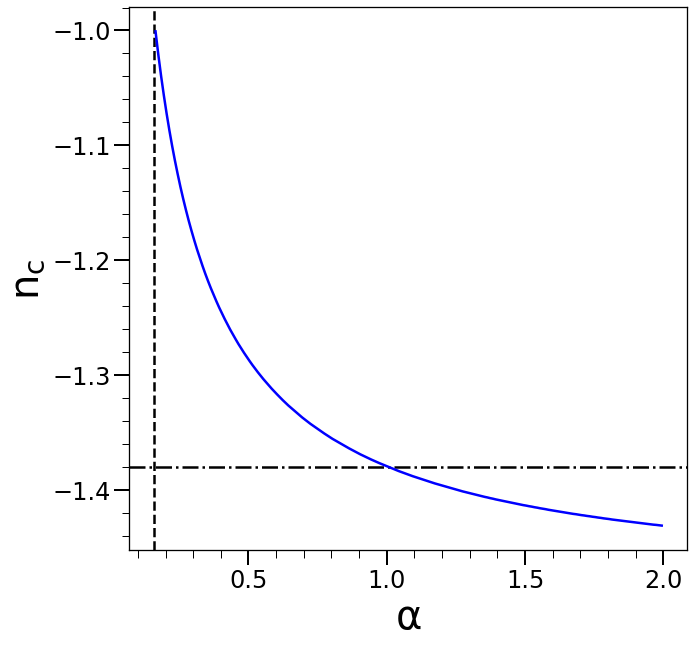}
    \caption{Left panel: coefficient $c$ characterizing the one-loop correction 
    to the PS in standard perturbation theory, as a function of $n$ in the 
    range $n<-1$ where the result is finite, for different values of 
    $\alpha$. Right panel: the critical value $n=n_c$ at which the 
    one-loop correction to the PS changes sign, as a function of $\alpha$. The black dash-dotted
      horizontal line corresponds to standard
    EdS for which $n \approx -1.38$, and the black dashed 
    vertical lines indicate the critical value $\alpha_c \approx 0.16$
    below which $c$ is always positive.  
    }\label{Figure-c-and-nc}
\end{figure*}
 This leads directly to the following expressions:
\begin{eqnarray}
      (\hat{M}_0+\hat{N}_0)&=&\frac{  2^{-(n+5)}\pi \left(n^2+2\right) \Gamma \left(\frac{1}{2}-n\right) \Gamma \left(\frac{n+1}{2}\right)}{\Gamma \left(2-\frac{n}{2}\right)^2 \Gamma
   \left(\frac{n+2}{2}\right)}, \nonumber \\
   \hat{M}_1&=&-\frac{2^{-(n+2)\pi} \Gamma \left(\frac{1}{2}-n\right) \Gamma \left(\frac{n+3}{2}\right)}{\Gamma \left(2-\frac{n}{2}\right)^2 \Gamma \left(\frac{n+2}{2}\right)},\nonumber \\
  \hat{M}_2&=&\frac{ 2^{-(n+2)\pi} \Gamma \left(\frac{1}{2}-n\right) \Gamma \left(\frac{n+3}{2}\right)}{\Gamma \left(2-\frac{n}{2}\right)^2 \Gamma \left(\frac{n+4}{2}\right)},\nonumber\\
   \hat{N}_1&=&\frac{3 \pi ^2 \csc \left(\frac{(n+3)\pi}{2}\right)}{8 \Gamma \left(1-\frac{n}{2}\right) \Gamma \left(\frac{n}{2}+4\right)},\nonumber\\
    \hat{N}_2&=&-\frac{3 \pi ^2 \csc \left(\frac{(n+3)\pi}{2} \right)}{16 \Gamma \left(2-\frac{n}{2}\right) \Gamma \left(\frac{n}{2}+3\right)}, \nonumber \\
\end{eqnarray}
from which it follows that
\begin{widetext}
\begin{eqnarray}
     c\Big(n,\alpha \Big)
     %&=& (\hat{M}_0+\hat{N}_0)+\frac{1+4\alpha}{1+6\alpha} \Big[\hat{M}_1 + \frac{1+4\alpha}{1+6\alpha} \hat{M}_2\Big] 
     % + \frac{1+2\alpha}{1+8\alpha} \Big[\hat{N}_1+ \frac{2\alpha}{1+6\alpha} \hat{N}_2\Big]\nonumber\\ 
     & = &\frac{2^{-(n+2)}\pi \Gamma \left(\frac{1}{2}-n\right)\Gamma \left(\frac{n+3}{2}\right)}{\Gamma \left(2-\frac{n}{2}\right)^2 \Gamma
   \left(\frac{n+2}{2}\right)}\Bigg[ \frac{\left(n^2+2\right)}{4(n+1)}+\Big(\frac{1+4\alpha}{1+6\alpha}\Big)
   \Bigg\{1+\Big(\frac{1+4\alpha}{1+6\alpha}\Big)\frac{1}{n+2}\Bigg\}\Bigg]\nonumber\\ 
   &&+\Big(\frac{1+2\alpha}{1+8\alpha}\Big)\frac{3\pi}{8}\Big(\frac{\Gamma \left(\frac{n+3}{2}\right)\Gamma \left(-\frac{n+1}{2}\right)}{\Gamma \left(1-\frac{n}{2}\right)\Gamma \left(\frac{n}{2}+3\right)}\Big)\Bigg[\frac{2}{n+6}-\Big(\frac{2\alpha}{1+6\alpha}\Big)\frac{1}{2-n}\Bigg].\nonumber\\
     \label{c-one-loop-deimnereg}
 \end{eqnarray}
\end{widetext}

Setting $\alpha=1$ in the individual expressions for $P_{13}$ and $P_{22}$ used to derive Eq.~(\ref{c-one-loop-deimnereg}), we have checked that we recover identical expressions to those  in \cite{pajer2013renormalization} and \cite{scoccimarro1996loop}.\footnote{As noted in \cite{pajer2013renormalization}, there is a sign error in one term in the expression for 
%\textcolor{green}{
the fourth term of $P_{22}$ given in \cite{scoccimarro1996loop}. The latter reference also defines a coefficient labeled $\alpha_\delta$ analogous to our $c$, but differing by a factor, with $\alpha_\delta= \frac{2}{\Gamma(\frac{n+3}{2})} c$.} 

A further check on the correctness of the expression Eq.~(\ref{c-one-loop-deimnereg}) is obtained by comparing with the exact result for the case $n=-2$ which, as detailed further below in Sec. \ref{UV divergences}, can be obtained directly using the expressions in Eq.~(\ref{Mhat-Nhat}) as  
\begin{equation}
\label{c_n=-2}
%c(n=-2, \alpha)= \frac{3 \pi ^2 (4 \alpha +1) (2 \alpha  (11 \alpha +5)+1)}{8 (6 \alpha +1)^2 (8 \alpha +1)}  
c(n=-2, \alpha)= \frac{3 \pi ^2 (4 \alpha +1) (22 \alpha^2 +10 \alpha+1)}{8 (6 \alpha +1)^2 (8 \alpha +1)}.
\end{equation}

The left panel of Fig. \ref{Figure-c-and-nc} shows $c(n,\alpha)$ as a function of $n$ for different chosen values of $\alpha$, including the canonical $\alpha=1$ case. Compared to the latter, the most evident
qualitative change as $\alpha$ varies is that the zero crossing of $c$, which is at $n=n_c\approx -1.38$ for $\alpha=1$, not only increases toward $n=-1$ as $\alpha$ decreases but actually ceases to exist at
a certain critical value of $\alpha$. The right panel of Fig. \ref{Figure-c-and-nc} shows the
quantitative behavior of $n_c$ as a function of $\alpha$. This critical value is none other than 
$\alpha_c$, the positive root of the function $f_{-1}$ discussed above, at which the leading 
divergence changes sign. Indeed  we can see this also by expanding our expression 
Eq.~(\ref{c-one-loop-deimnereg}) around $n \rightarrow -1$, where it has a simple pole,
which gives
\begin{eqnarray}
   c(n=-1+\xi, \alpha)&=&-\frac{(7-14\alpha-176\alpha^{2})}{15 (6 \alpha +1) (8 \alpha +1) \xi }\nonumber\\
   & +&\frac{4 (2 \alpha +1) (4 \alpha +1)}{9 (6 \alpha +1)^2}+\cdots
   \label{cd-at-n-eq-min1} 
\end{eqnarray}

We note also that, other than very close to the divergence, $c$ is a very slowly varying function of $\alpha$ in the range of $\alpha$ which is relevant to current standard type models, for which the logarithmic linear growth rate varies between $\alpha=1$ (and high redshift) and $\alpha \sim 0.5$. As discussed in
P1, the correction to the one loop PS relative to the EdS value in these models can be well approximated (to about $20-25 \%$) by calculating in a gEdS model with an effective value at $z=0$ of $\alpha \sim 0.9$ (which represents an appropriately averaged growth rate over the cosmological evolution). 

\section{Numerical tests of predicted $\alpha$-dependence (for $n=-2$)}  
\label{Numerical tests} 

\begin{figure}[t]
\centering\includegraphics[width=7cm, height=7cm]{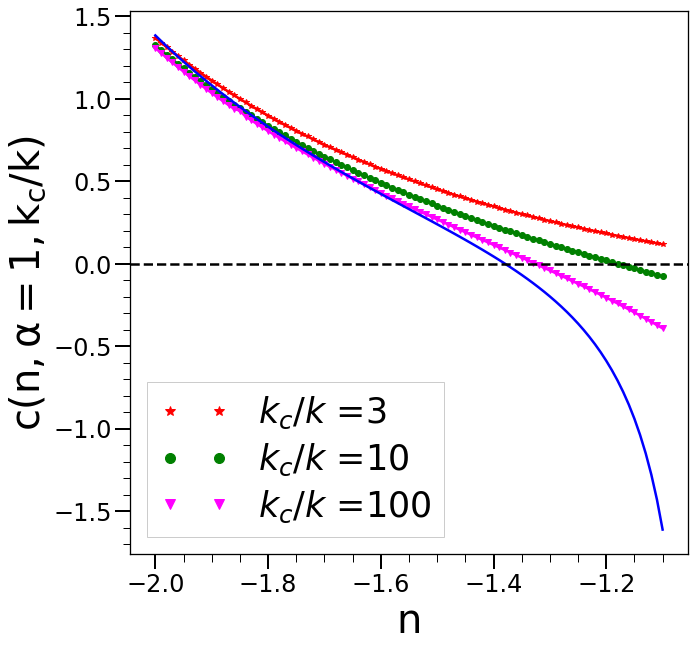}
\caption{Numerically evaluated $c(n,\alpha=1, k_c/k$) for different indicated values of 
the cutoff $k_c$, as a function of $n$. Also shown is the exact result (solid line) 
obtained using dimensional regularization. }
\label{Fig-NumInt-c-different-cutoffs}
\end{figure}

\begin{figure}[t]
\centering\includegraphics[width=7cm, height=7cm]{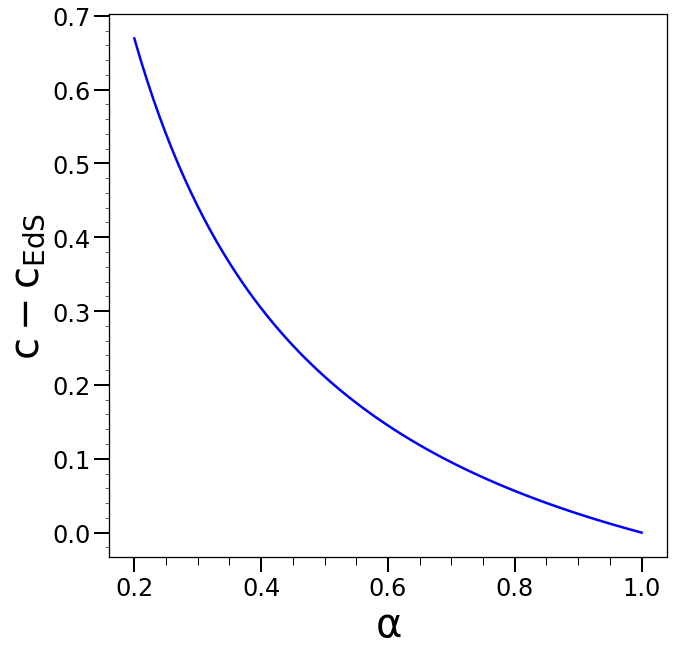}
\caption{Difference between the coefficient $c(n=-2,\alpha)$ and its value 
in the standard EdS model $c_{\rm EdS}=c(n=-2,\alpha=1)$, for $\alpha$ in the 
range explored by our suite of simulations. Note that it is this
quantity multiplied by $\Delta_L^2$ which gives the fractional
change in the predicted PS, and one loop PT is expected only to
apply for $\Delta_L^2 \ll 1$. At $\Delta_L^2 \sim 0.1$ the 
predicted maximal change in power, for $\alpha=0.25$, is
thus of order of $5\%$. This can be compared with the 
much smaller changes in standard (LCDM-like) models,  
of order $0.5 \%$ at $z=0$ (see P1 and 
\cite{takahashi2008third,Garny2022Twoloop,fasiello2022perturbation})
%$0.2$ to $1.0$   and   $c_{EdS}= c(n,\alpha=1)$.
}
\label{c-cEds}
\end{figure}

In this section we compare the results of numerical simulations with the analytical result 
given by Eq.~(\ref{c-one-loop-deimnereg}). While it is potentially of interest to consider 
a wide range of different $n$ and $\alpha$, we limit ourselves here to probing the
$\alpha$-dependence (which is the novelty of our analysis) of the result for $n$
in the regime where we expect that this result may actually provide a good approximation
i.e. where the ultraviolet sensitivity of the result is weak, for $n$ well below $-1$.
To quantify this a little more we show, in Fig.~\ref{Fig-NumInt-c-different-cutoffs},  
the results of a determination of $c(n,\alpha, k_c/k)$ by direct numerical integration 
for the different indicated values of the cutoff $k_c/k$. The ultraviolet sensitivity 
as expected diminishes markedly as $n$ decreases. In the left panel of Fig.~\ref{Figure-c-and-nc}
we see, on the other hand, that the $\alpha$-dependence remains quite uniform for 
$n<-1.5$. If $n$ decreases too close to $n=-3$, however, the dynamical
range of a simulation due to the finite simulation box size will become
very limited. We thus consider the value $n=-2$. Figure \ref{c-cEds} shows, 
for this value of $n$, the predicted difference as a function of $\alpha$
between the coefficient $c$ and its value for $\alpha=1$. Given that the modification of the PS is 
proportional to $c$ multiplied by $\Delta^2_L(k)$, and that the one-loop calculation is expected
to be valid only for small values of the latter, it is evidently of interest to 
simulate smaller values of $\alpha$ for which the difference in power is amplified. 
We consider here simulations with $N=256^3$ particles, and 
the values $\alpha=1.00,0.7,0.5,0.33,0.25$. 
The lower limit $\alpha=0.25$ is imposed, as we will explain further below, 
because the numerical cost of the simulations increases strongly as $\alpha$ decreases. 
Nevertheless this value is sufficient to give predicted 
changes in the power of order $5\%$ for $\Delta_L^2=0.1$, much greater (and 
therefore much easier to measure numerically) than the predicted 
changes of $\sim 0.5 \%$ in standard (LCDM-like) models 
(see \cite{takahashi2008third,Garny2022Twoloop,fasiello2022perturbation}). As discussed in P1, the latter can be well 
approximated by using a gEdS model with $\alpha \approx 0.9$.

\subsection{Simulation method}

\begin{figure*}[t]
    \includegraphics[width=8cm,height=8cm]{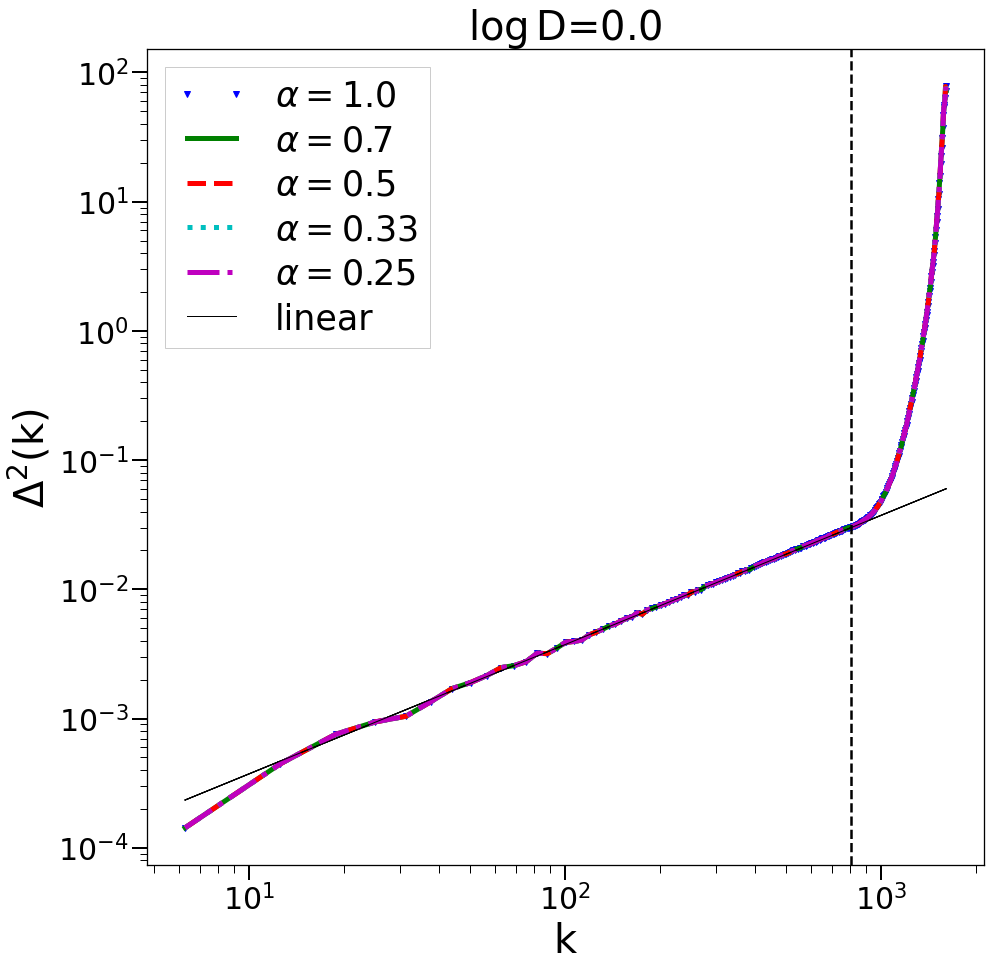}
    \includegraphics[width=8cm,height=8cm]{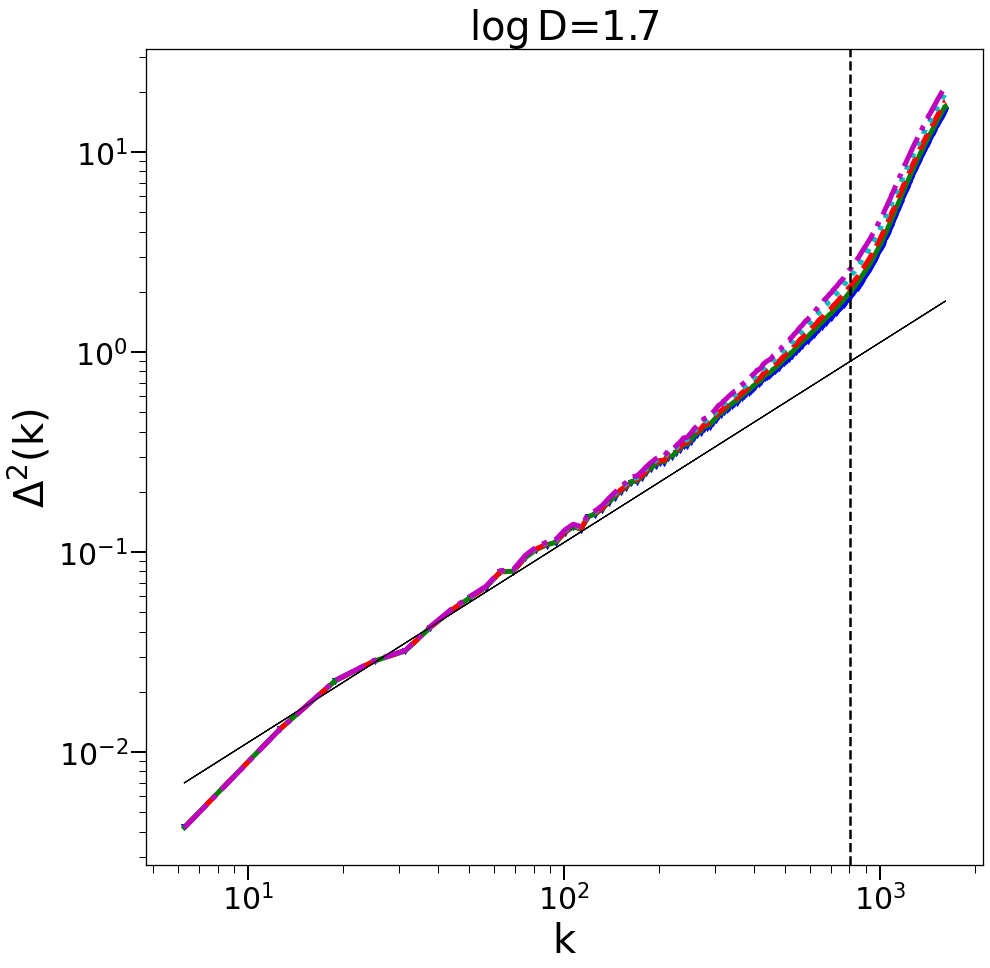}
    \\[\smallskipamount]
    \includegraphics[width=8cm,height=8cm]{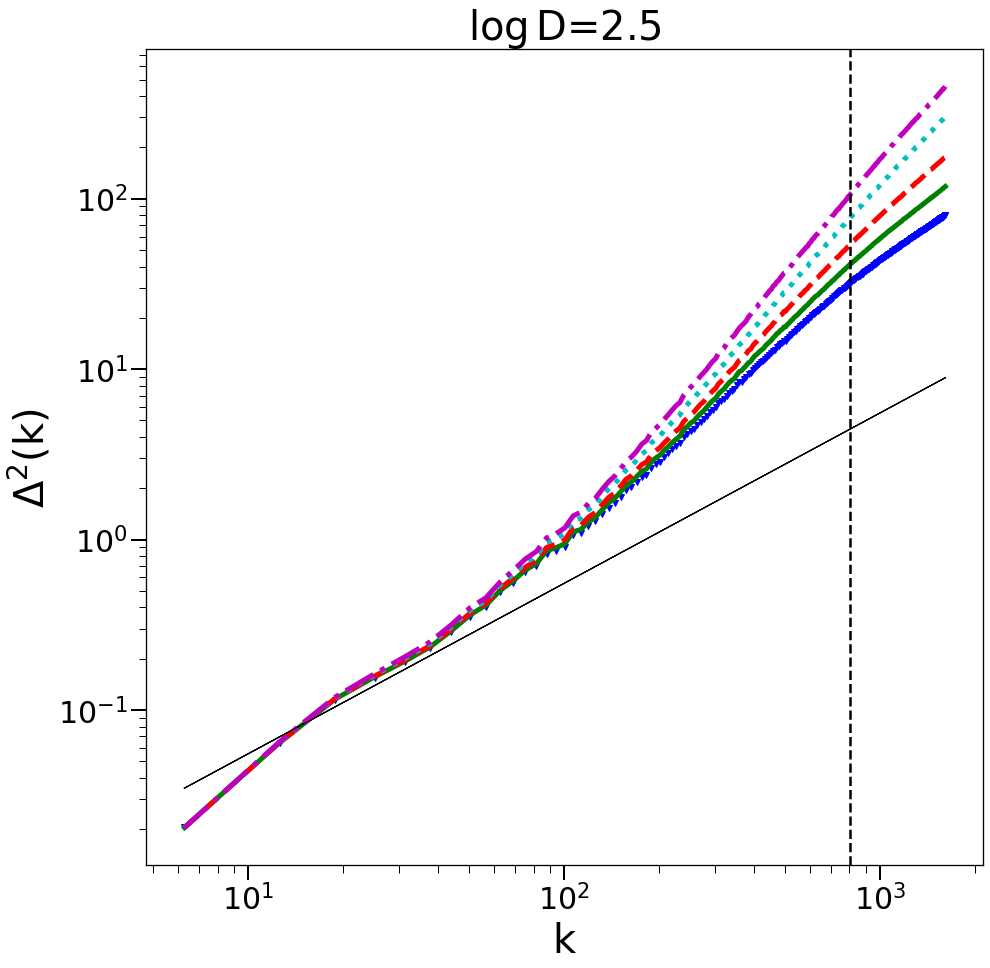}
    \includegraphics[width=8cm,height=8cm]{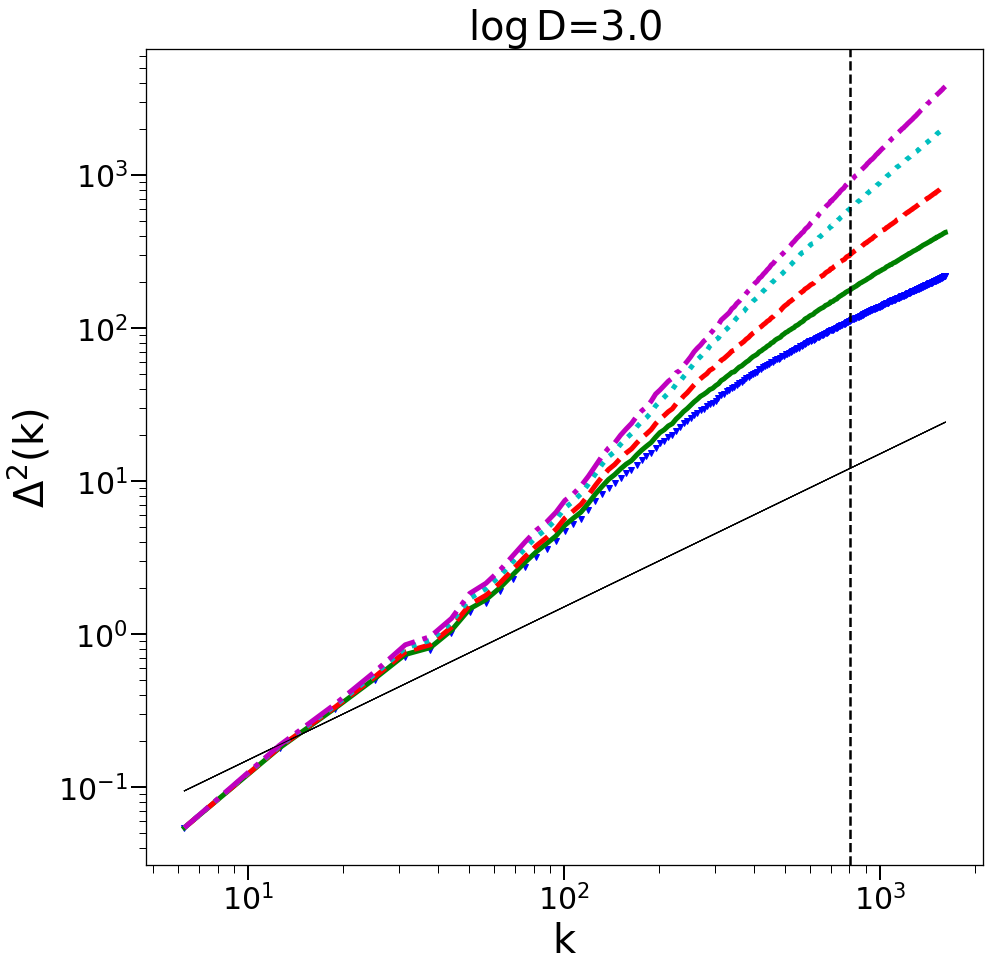}
    \caption{
    Dimensionless PS measured in our suite of five simulations, with the indicated values of $\alpha$, as a function of $k$ (in units in which the box size $L=1$). The solid black line is the  dimensionless linear PS $\Delta_{L}^2(k)$. The first panel is the initial configuration (with identical power in each simulation) and the other three progressively more evolved snapshots.  \label{DeltaSq_measured_4times}}
\end{figure*}

\begin{figure*}[t]
    \includegraphics[width=8cm,height=8cm]{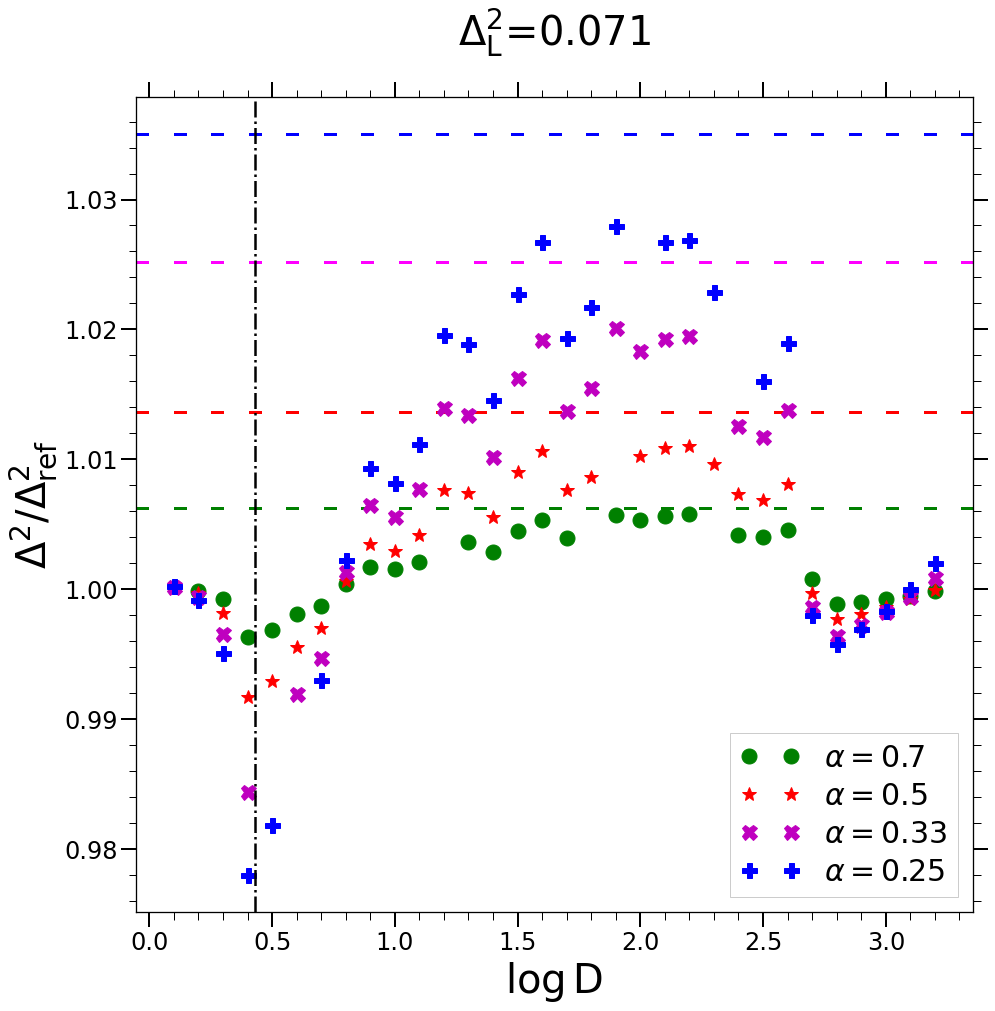}
    \includegraphics[width=8cm,height=8cm]{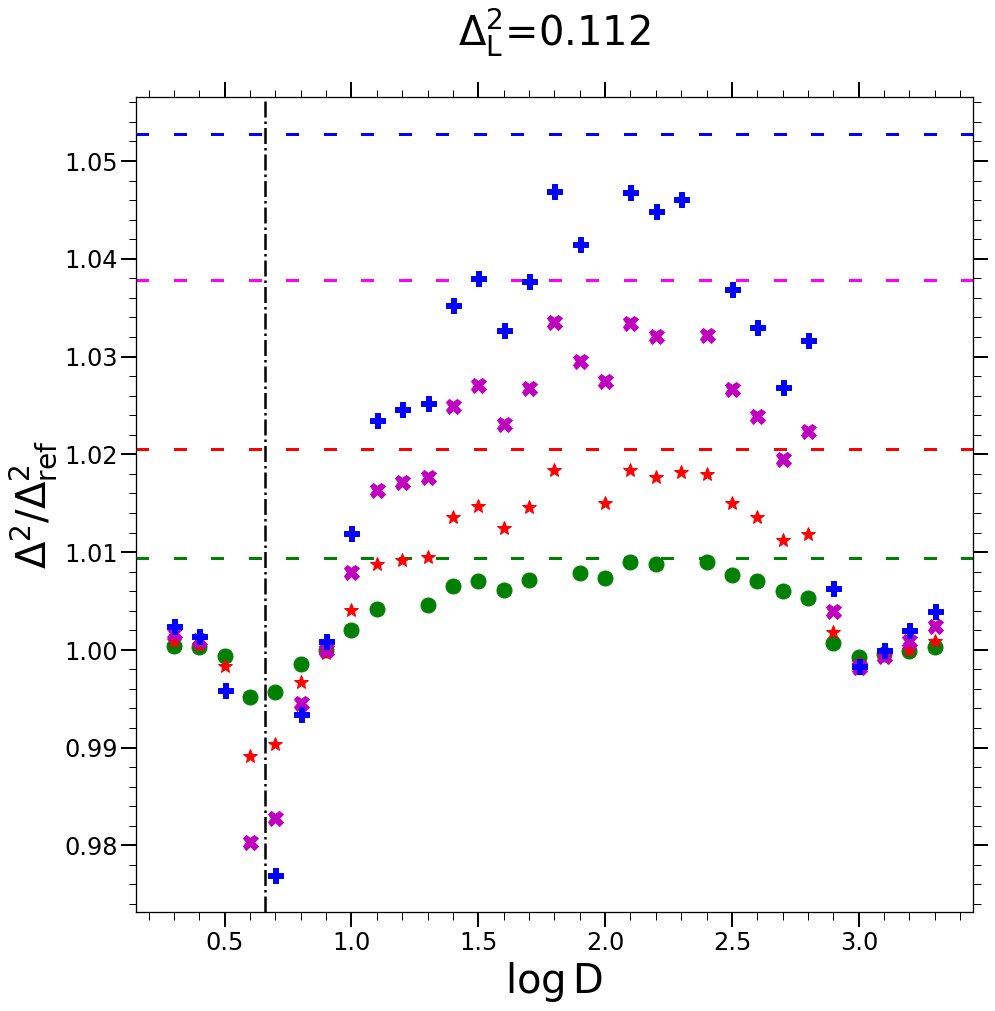}
    \\[\smallskipamount]
    \includegraphics[width=8cm,height=8cm]{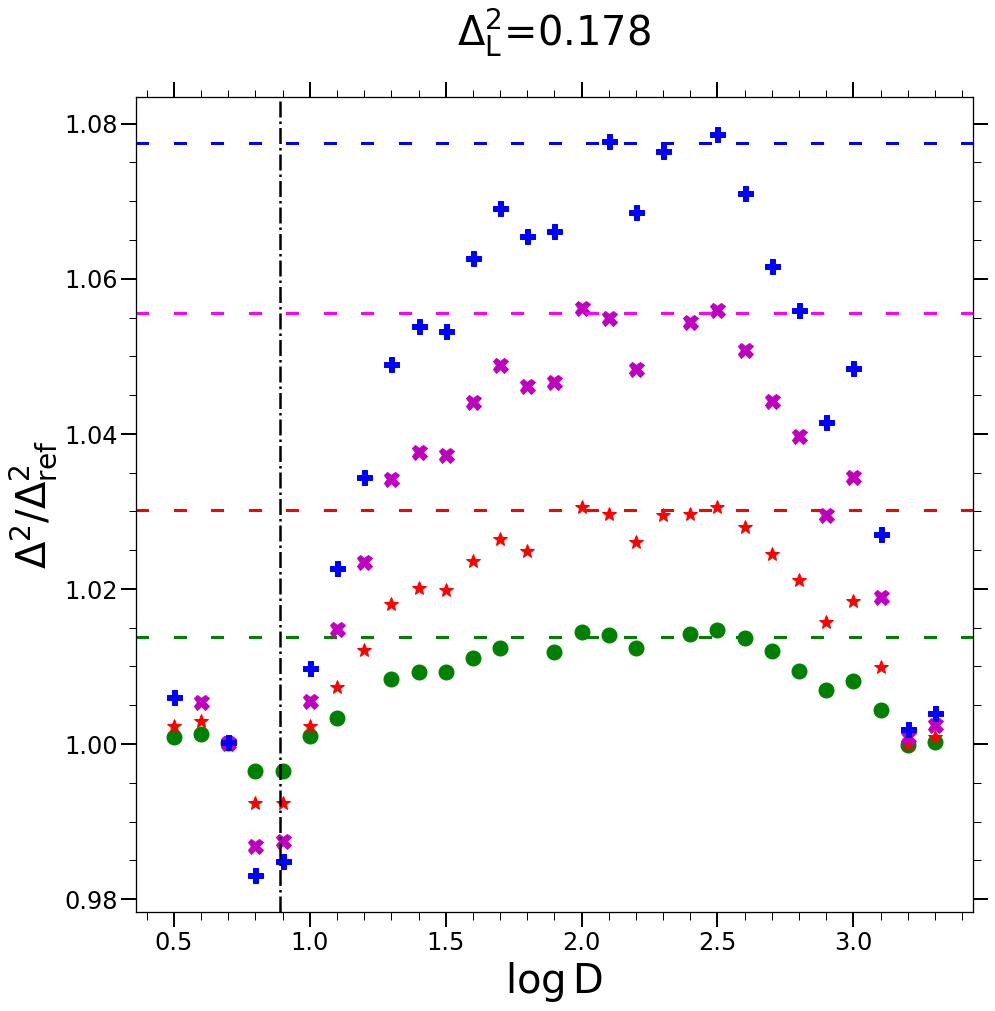}
    \includegraphics[width=8cm,height=8cm]{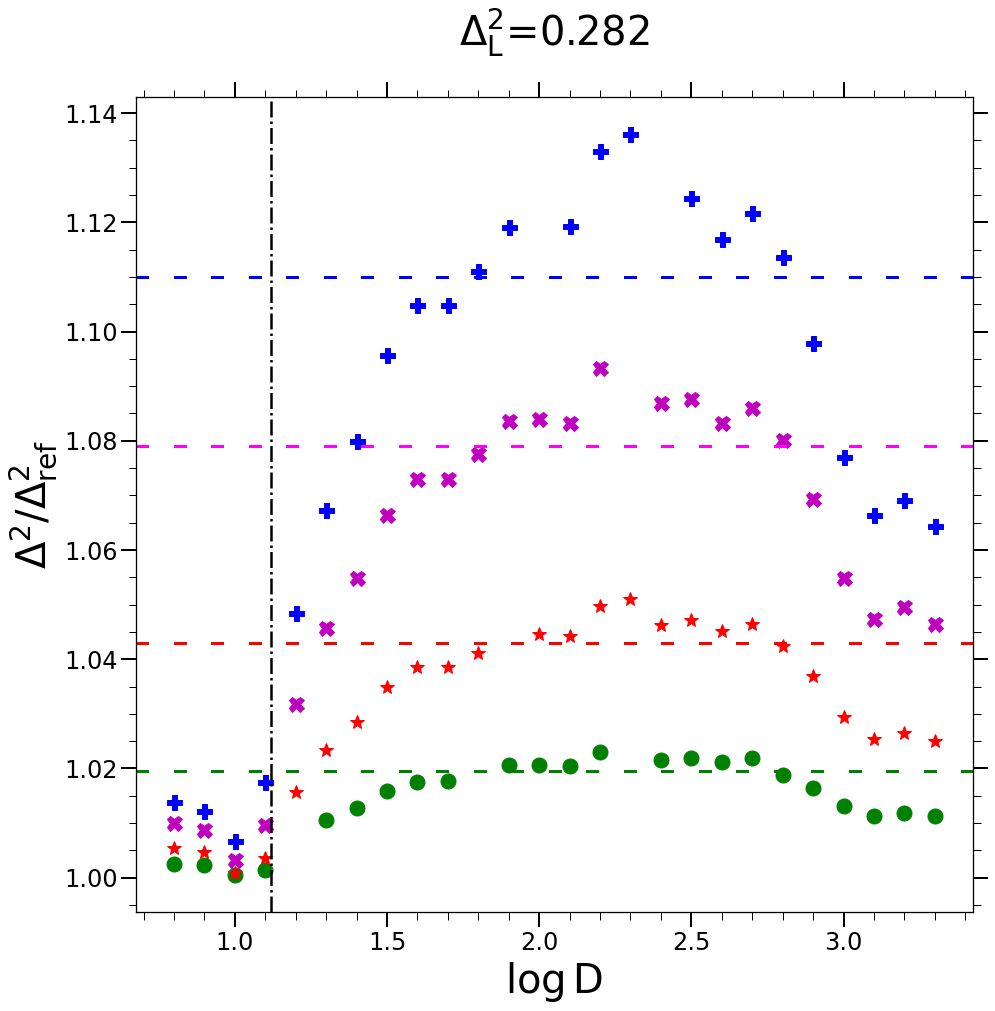}
    \caption{Ratios of PS measured in the four simulations with $\alpha <1$
    to that measured in the standard EdS ($\alpha=1$) simulation, as a 
    function of time parametrized as $\log D$. Each 
    plot corresponds to the indicated chosen value of the theoretical input 
    dimensionless PS $\Delta_L^2$. Self-similar behavior (i.e. a result
    independent of the scales introduced by the $N$-body simulation) corresponds
    to a constant value. The different horizontal lines correspond to the
    (self-similar) ratios predicted by one loop standard perturbation theory.
    The vertical line on each plot indicates the time at which $k=k_N$, the Nyquist 
    wave number of the particle grid. }
   \label{RelativePower_measured}
\end{figure*}
Our numerical results here have been obtained using $N$-body simulations performed with an appropriately modified version of the GADGET2 code \cite{gadget} 
as described in detail in \cite{Benhaiem}, and further in \cite{dbenhaiem_PhD}. 
Indeed the class of scale-free models we are considering cannot be simulated by the standard version of 
GADGET2 code, which allows only expanding  backgrounds specified  by the standard cosmological parameters. 
The gEdS cosmology has been implemented instead by modifying the module of the GADGET2 code which allows simulation also of a {\it static} universe (i.e. of an infinite periodic system without 
expansion). The usual equations solved in $N$-body simulations for particles in an expanding background are given in comoving 
coordinates $\mathbf{x}$ as
\begin{equation}
	\frac{d^2 {\bf x}_i}{dt^2} +
	2H \frac{d{\bf x}_i}{dt}  = \frac{1}{a^3} \mathbf{ F} _i
	\label{3d-equations-1}
\end{equation}
where the gravitational force is
\begin{equation}
	\mathbf{F}_i  =- Gm \sum_{j \neq i}^P
	\frac{{\bf x}_i - {\bf x}_j}{\vert {\bf x}_i - {\bf x}_j \vert^3} 
	W_\varepsilon (\vert {\bf x}_i - {\bf x}_j \vert) \,
	\label{3d-equations-2}
\end{equation}
with $W_\varepsilon$ a function that smooths the singularity of
the Newtonian force at zero separation, at a characteristic scale 
$\varepsilon$, and the ``P'' in the sum indicates that there is
a sum over the copies of the periodic system. 
As discussed in further detail in \cite{Benhaiem}, these equations
can be recast, by the simple change of time coordinate $\tau=\int dt \, a^{-3/2}$,
as
\begin{equation}
\frac{d^2 {\bf x}_i}{d\tau^2} +
	\Gamma \frac{d{\bf x}_i}{d\tau}  =  {\bf F}_i \,,
\end{equation}
where 
\begin{equation}
\Gamma=\frac{1}{2}a^{3/2} H=\frac{1}{2}a^{-1}\frac{da}{d\tau}\,.
\end{equation}
Thus the equations of motion are just those of self-gravitating particles in a nonexpanding system 
subject to a simple fluid damping. The family of $gEdS$ cosmologies corresponds to models given 
by a constant value of $\Gamma$, with
\begin{equation}
\Gamma=\kappa \sqrt{2\pi G \rho_0 /3}%=\frac{\kappa}{3t_0}\,,
\end{equation}
where $\rho_0$ is the mean mass density at some chosen reference time.
The static universe module of GADGET2 has thus been modified to include this constant fluid-damping term, 
keeping the original ``kick-drift-kick'' structure of its leap-frog algorithm and modifying appropriately the ``kick'' and ``drift'' operations. The structure of the code is otherwise unchanged. Further 
details and various tests of the modified code have been described in \cite{Benhaiem}, in particular tests of energy conservation (using the so-called Layzer-Irvine equations) as well as a direct comparison showing excellent agreement between simulations of the standard EdS (i.e. $\alpha=1$) model using the existing 
GADGET2 expanding universe module and the new modified static universe module.

To generate initial conditions we use the canonical method, applying displacements to the simulation particles 
initially placed on a perfect lattice, and ascribing corresponding initial velocities, as prescribed by the 
Zeldovich approximation, for a random realization of a Gaussian fluctuation field with the chosen input PS
(for more details  see e.g. \cite{bertschinger1995cosmics} and \cite{joyce2007quantification}). 
At the starting time, $a_0$, the initial amplitude of the PS has been 
set using the specific choice (following the criteria of \cite{jain1998self, knollmann2008dark})
%$A_0$ are given by 
\begin{equation}
       \Delta_{L}^2(k_N,a_0)=0.03 %\frac{A_0 D_0^2 k_N^{3+n}}{2\pi^2}=0.03,
\end{equation}
%where $D_0=1$ is the growth factor, $n=-2$, and 
where $k_N$ is the Nyquist frequency of the initial particle grid. We use the same realization of 
the initial density field in all five simulations. The initial displacements are thus
identical in the five simulations, and the initial velocities simply rescaled 
appropriately for each $\alpha$ (since the Zeldovich displacement is proportional to $D$).

Outputs of the simulations have been saved, starting from the initial time, at times defined by
\begin{equation}
%t_s=\log{\frac{D(a)}{D(a_0)}}= 0.1 s 
t_s=\log D(a)= 0.1 \,s 
\end{equation}
where $D(a)$ is the linear growth factor, defined so that $D(a_0)=1$, 
and $s=0,1,2 \cdots 33$. Thus the predicted linear power spectrum in each 
simulation is identical at each output, and the final output, at  $a=a_f$,
corresponds to an amplitude $\Delta_{L}^2(k_b,a_0)=e^{6.6} 0.03/256 \approx 0.17$ 
at the fundamental mode $k_{b}=2\pi/L$ of the periodic box. As we will see
below by this time the finite box size corrections are very dominant over the
very small effects we are seeking to measure (at the few percent levels).

To calculate the power spectrum based on data from $N$-body simulations, we have used the publicly available POWMES code \cite{Colombi} with the size of FFT grid equal to $512^3$ (compared to the $256^3$ initial particle grid) and without any  ``foldings." This is quite sufficient resolution for the analysis here, focusing on smaller $k$. 
 
\subsection{Results}

Figure \ref{DeltaSq_measured_4times} shows the dimensionless PS measured in the five simulations,
at the starting time and at three subsequent times. Also shown (solid black line) is the linearly evolved theoretical input PS (which, by construction, is the same at each time for all the simulations). Likewise we see that the initial PS of the IC is identical at the starting time.    
Inspecting the $\alpha$ dependence of the evolving PS, 
we observe 
%in Fig. \ref{DeltaSq_measured_4times} 
a qualitative behavior in line 
with Fig.~\ref{c-cEds}: as $\alpha$ decreases the nonlinear power increases. However 
this trend with $\alpha$ is in fact clearly
visible in these plots only starting from $\Delta^2$ approaching
unity, where we do not expect perturbation theory to apply. Indeed as we have discussed,
Fig. \ref{c-cEds} implies changes to the nonlinear power of at most about
ten percent. The origin of the amplification of the highly nonlinear power we 
observe in this plot --- and more particularly the steepening of its slope as a function 
of $\alpha$ has been discussed at length in 
%\textcolor{green}{
\cite{benhaiem2014self}. Here 
we focus instead on the perturbative regime.

We also see in Fig. \ref{DeltaSq_measured_4times} the visible effects of finite mode sampling on the small $k$ modes (i.e. small $\Delta_L^2$) which are relevant for the regime we are interested in: indeed for smaller $k$ 
there are clearly, at the initial time, visible fluctuations 
of the measured PS $\Delta_{sim}^2(k)$ relative to the theoretical linear PS power spectrum $\Delta_L^2(k)$.\footnote{Note that the fundamental mode in our units
is $2\pi$. The visible ``dip''
at small $k$ arises from just the
first sparsely populated bin.}
%in the realization of the initial conditions in the periodic box. 
Thus we expect that a comparison of the observed power with the theoretical prediction can be accurate at best up to a systematic error of order
$\delta=(\Delta^2(k)/\Delta_{sim}^2(k)) -1$, while if we 
consider the measured {\it ratio} of the power between two simulations
we can expect accuracy instead of $\delta \times [c(n,\alpha)-c(n,\alpha=1)]$. 
In order to measure the very small effects predicted, we therefore
consider this relative measurement, using (arbitrarily) 
$\alpha=1$ as our reference. 

Figure \ref{RelativePower_measured} shows results for the ratio of the 
PS measured in the four simulations with $\alpha<1$ to that in the standard
EdS case. Following the analysis method developed in %\textcolor{green{
\cite{joyce2021quantifying,maleubre2022accuracy},
each panel is for a different bin of
$\Delta_L^2$ (corresponding to a fixed bin of 
rescaled wave number $k/k_{NL}$), and shows
the ratios measured in the different snapshots. 
The indicated values
of $\Delta_L^2$ correspond to those
calculated for the theoretical input 
PS spectrum at the geometric
center (in $k$) of the bins, 
which are equally spaced in log
space with $\Delta \log_{10} k=0.1$.
We underline that, because the points are plotted as a function of $\log D$, the differences measured in these plots arise purely from the nonlinear evolution.
Further the measured power spectrum is self-similar 
if and only if it is a function of $\Delta_L^2(k)$ only i.e. if it is constant in each plot.
The ratios of the measured (self-similar) power predicted by 
eq.~(\ref{c-one-loop-deimnereg}) for each value of $\alpha$, 
is indicated by a horizontal line.\footnote{The finite size of the bins
has also been taken into account in this latter calculation 
but only very marginally modifies the result.}

The behavior we observe in the plots in Fig. \ref{RelativePower_measured} is qualitatively similar 
to that in analogous plots from the (much larger, but standard EdS) simulations
analyzed and discussed in \cite{maleubre2022accuracy}.
The points from any given simulation, at the chosen rescaled wavenumber $k/k_{NL}$ in each plot, display approximately, in differing 
degrees and ranges of time, the flat 
behavior corresponding to self-similarity.   
The strong temporal evolution at early times 
arises from the ultraviolet cutoffs (grid spacing, force smoothing), while the 
strong suppression at later times arises
from the finite box size. Indeed the latter
sets in at later times in the successive 
plots, as $\Delta_L^2$, and therefore the associated $k$ at a given time, increases.
The vertical line in each plot indicates
the time at which $k$ corresponds 
to the Nyquist wave number of the 
initial grid, which likewise increases
as $\Delta_L^2$ does so.  In the upper
two plots the results are also, because 
they correspond to smaller $k$ at any 
time, significantly more noisy. The
plateaus can just about be discerned
within a large approximate error bar 
given by the amplitude of the scatter
in the flattest five or six points.
Comparing these plateau values with 
the predicted ones (given by the dotted
lines) we see that the overall agreement 
is very good, and most particularly in 
the cases where the plateau is very well
defined, notably in the lower two plots.
It appears that the theoretical
value is systematically a little high in 
the first two plots. This can be attributed
to the fact that this theoretical prediction is calculated with the theoretical input
$\Delta_L^2$, which fluctuates more at these smaller $k$ relative to the 
actual initial conditions. The apparently
slightly low theoretical values for
the smallest $\alpha$ simulation 
in the last plot probably reflect 
the increasing contribution of 
higher order corrections
expected as the amplitude of the deviations
grow (in these cases above about ten percent).
We conclude thus that the $\alpha$-dependence of the PS observed in our
simulations are apparently in good agreement 
with the one-loop PT predictions.

\section{$\alpha$ dependence of UV divergences and their regulation} 
\label{UV divergences} 

\begin{table*}
\caption{ Analytical expressions for the six integrals, $\hat{M}_{i}$ and $\hat{N_i}$ (for $i=0,1,2$), 
for $n=1$, $n=0$, $n=-1$ and $n=-2$, up to linear order in $k/k_c$ and in the limit $\epsilon=0$.
}
\begin{ruledtabular}
\begin{tabular}{ccccc} 
$\hat{M}_i$, $\hat{N}_i$   &   n=1 & n=0  &   n=-1 & n=-2 \\ 
 \hline
 $\hat{M}_0$ &  $\frac{1}{2}\big(\frac{k_c}{ k}\big)-\frac{7}{8}$& $\frac{\pi ^2}{16}$ &  $\frac{1}{3} \log \big(\frac{k}{\varepsilon }\big)$ &  $\frac{k}{3 \varepsilon }$ \\ 
 $\hat{M}_1$ &  $-\frac{4}{3}\big(\frac{k_c}{k}\big)+\frac{9}{4}$ & $-\frac{\pi ^2}{8}$ &  $-\frac{4}{9}$ &$0$  \\ 
 $\hat{M}_2$ &  $\frac{16}{15}\big(\frac{k_c}{k}\big)-\frac{3}{2}$ & $\frac{\pi ^2}{8}$  &  $\frac{8}{9}$ & $\frac{3 \pi ^2}{16}$ \\ 
 $\hat{N}_0$ &  $-\frac{1}{6}\big(\frac{k_c^2}{ k^2}\big)$ & $-\frac{1}{3}\big(\frac{k_c}{ k}\big)$ &  $\frac{1}{3} \log \big(\frac{\varepsilon }{k_c}\big)$ & $-\frac{ k}{3 \varepsilon }$\\ 
 $\hat{N}_1$ & $\frac{2}{5}\big(\frac{k_c^2}{ k^2}\big)+\frac{4}{35} \log \left(\frac{4k}{k_c}\right)-\frac{958}{1225}$ & $\frac{4}{5}\big(\frac{k_c}{ k}\big)-\frac{\pi ^2}{16}$  &  $-\frac{4}{5} \log \big(\frac{k}{k_c}\big)+\frac{32}{75}$ & $\frac{3\pi^2}{16} $  \\ 
 $\hat{N}_2$ &  $-\frac{2}{3}\big(\frac{k_c^2}{ k^2}\big)-\frac{4}{5} \log \left(\frac{4k}{k_c}\right)+\frac{142}{75}$ & $-\frac{4}{3}\big(\frac{k_c}{ k}\big)+\frac{3 \pi ^2}{16}$   &  $\frac{4}{3} \log \big(\frac{k}{k_c}\big)$ & $-\frac{3 \pi ^2}{16}$ \\ 
 %[2ex] 
 %\hline
\end{tabular}
\label{Table-MandN-integer-n}
\end{ruledtabular}
\end{table*}

\begin{table*}
\caption{Analytical expressions for the one-loop coefficients $c(n,\alpha, k_c/k)$
 for $n=1$, $n=0$, $n=-1$ and $n=-2$ up to linear order in $k/k_c$ and in the limit $\epsilon=0$. }
\begin{tabular}{|c| c|} 
 \hline
  n  &   $c(n, \alpha, k_c/k)$ \\ 
 \hline
 \hline
 1  &   $\frac{(7-14\alpha-176\alpha^{2})}{30(6\alpha+1)(8\alpha+1)}(\frac{k_{c}}{k})^2       +\frac{(7+36\alpha+92\alpha^{2})}{30(1+6\alpha)^{2}} \frac{k_{c}}{k}+\frac{4 (2 \alpha +1) (8 \alpha -1) \log \left(\frac{k_c}{4k}\right)}{35 (6 \alpha +1) (8 \alpha +1)}$\\
 &$-\frac{4 \alpha  (\alpha  (168072 \alpha +133249)+59990)+26667}{29400 (6 \alpha +1)^2 (8 \alpha +1)}$ \\
 0  &   $\frac{(7-14\alpha-176\alpha^{2})}{15(6\alpha+1)(8\alpha+1)}\frac{k_{c}}{k}+\frac{\pi ^2 \alpha  (4 \alpha +1) (5 \alpha +1)}{2 (6 \alpha +1)^2 (8 \alpha +1)}$  \\
 -1 &   $\frac{(7-14\alpha-176\alpha^{2})}{15 (6 \alpha +1) (8 \alpha +1)}\ln\frac{k_c}{k}+\frac{4 (2 \alpha +1) (32 \alpha +7) (52 \alpha +7)}{225 (6 \alpha +1)^2 (8 \alpha +1)}$  \\
 -2  &   $\frac{3 \pi ^2 (4 \alpha +1) (2 \alpha  (11 \alpha +5)+1)}{8 (6 \alpha +1)^2 (8 \alpha +1)}$   \\
% [2ex] 
\hline
\end{tabular}
\label{Table-c-integer-n}
\end{table*}

General considerations (see e.g. \cite{Peebles1980LSS}) lead one to expect that nonlinear cosmological clustering should be ultraviolet insensitive for power-law initial conditions $P(k) \propto k^n$ provided 
$n<4$. For scale-free models, with an EdS expansion law, such cutoff independence implies self-similarity. Numerous
numerical studies confirm that such self-similarity is indeed observed, with different authors exploring different 
ranges of $n$, up to $n=2$ (see e.g.
\cite{efstathiou1988gravitational,padmanabhan1995pattern,Colombi1996Self,jain1996self,jain1998self,Bottaccio2002Clustering,smith2003stable,Baertschiger2007aGravitational,Baertschiger2007bGravitational,Baertschiger2008Gravitational, Orban2011Self,benhaiem2014self}). The ultraviolet divergences
which render the predictions of standard perturbation theory (SPT) undefined as $n$ approaches $-1$ from below 
are a priori therefore unphysical. The PS of standard cosmologies, however, at 
large $k$ has a behavior (typically $\sim k^{-3} \log k$) which leads to 
finite SPT predictions. Nevertheless there is also a region where the effective 
logarithmic slope of the PS corresponds to that of the ultraviolet divergent 
region, and one expects then that the associated unphysical divergences lead
to inaccuracies of the predictions of SPT. In the last number of years 
there has been much interest and work on the so-called effective field
theory (EFT) approach to the regulation of this ultraviolet 
divergences (see e.g. \cite{Baumann2012Cosmo,carrasco2012effective,pajer2013renormalization,hertzberg2014effective,mercolli2014velocity,porto2014lagrangian,carroll2014consistent,carrasco2014effective,carrasco20142,senatore2014redshift,senatore2015ir,baldauf2015bispectrum,vlah2015lagrangian,Angulo2015OneLoop,Baldauf2015EFTTwoLoop,foreman2016eft,steele2021precise}). This theory provides a systematic approach to the problem 
directly inspired from that used in high energy physics. 

Without employing the full machinery of EFT, we can recover very simply its
results for the class of model we are considering. To do so we impose 
a finite ultraviolet cutoff in the PS (i.e. we take $k_c$ to be finite)
above, and then consider how $k_c$ can scale with $k$ in a manner
compatible with self-similarity. 

For $n \geq -1$,  an analytical expression for $c(n, \alpha, k_c/k)$ (with $k_c/k$ finite)
can be found for integer values of $n$. To do so, as shown e.g. by \cite{makino1992}, 
one can conveniently rewrite the $\hat{M}_i$ double integrals by breaking up 
the integration range as 
   \begin{eqnarray}\label{Integral-limit-P22}
    \int_{\epsilon}^{1/\omega} dr  \int_{\mu_{min}}^{\mu_{max}}d\mu&=&
    \int_{\epsilon}^{1-\epsilon}  dr \int_{-1}^{1}d\mu\nonumber\\ 
    & &+\int_{1+\epsilon}^{(1/\omega)-1}dr
    \int_{-1}^{1}d\mu \nonumber\\ 
    & &+\int_{1-\epsilon}^{1+\epsilon}  dr\int_{-1}^{(1+r^2-\epsilon^{2})/2r}d\mu
    \nonumber\\ 
    & &+\int_{(1/\omega)-1}^{1/\omega}  dr\int_{(1+r^2-\omega^{-2})/2r}^{1}d\mu,\nonumber
\end{eqnarray}
where  $\epsilon=(\varepsilon/k)$ and $\omega=(k/k_c)$. For the $\hat{N}_i$ integrals 
we simply divide the integration range over $r$ into 
$\epsilon$ to 1 and from 1 to $1/\omega$. For each of the resulting integrals
an explicit analytic expression can be obtained (using $Mathematica$ \cite{Mathematica}), 
and written as a series expansion about $\epsilon=0$ or $\omega=0$, with poles 
associated with the divergences we have analyzed. As previously discussed
the divergences as $\epsilon \rightarrow 0$ in $\hat{M}_0$ and $\hat{N}_0$
cancel for $n>-3$.   The results for the individual integrals, $\hat{M}_i$ and $\hat{N}_i$,  
are shown in Table~\ref{Table-MandN-integer-n} and the resulting expressions for 
$c(n, \alpha, k_c/k)$ in Table~\ref{Table-c-integer-n}. In each case 
we have included terms in the expansion around $\omega=0$ and 
$\epsilon=0$ which do not vanish when the latter goes to zero. 
We note that the terms which diverge as $k_c/k$ are in agreement
with the results for the leading ultraviolet divergences given 
in Sec. \ref{PS-in-GSF}, with the leading divergence $\sim k_c^{n+1}$ 
and the following one at $\sim k_c^{2n-1}$. Further in the expressions
for $c$ we recover exactly the factors proportional 
to the $\alpha$-dependent coefficients $f_{-1}$, $f_{1/2}$ and $f_{1}$
given in Eqs.~(\ref{UV-factors})-(\ref{UV-factor1/2}) and (\ref{UV-factor-f1}).

In order to respect self-similarity it is sufficient to choose a regularization 
$k_c/k$ which is assumed to be some function of $\Delta_L^2(k)$. The simplest
and natural choice is to take
\begin{equation}
k_c \propto k_{NL}    
\end{equation}
i.e. to assume that the effective cutoff in the one-loop integrals is set by
the nonlinearity scale. Using this prescription we write the regularized
result first as 
\begin{eqnarray}
\label{c-regulated-n>-1}
\tilde{c}_{reg} (n,\alpha, \frac{k_{NL}}{k} ) &=&c \Big(n,\alpha, \frac{k_c}{k}=\gamma \frac{k_{NL}}{k} \Big) \nonumber \\
       &=& \lim_{\lambda \rightarrow \infty} \Bigg[c \Big(n,\alpha, \frac{k_c}{k}=\lambda \Big)
       + \Delta c \Big(n,\alpha, \gamma, \lambda \Big)\Bigg] \nonumber
       %+ \Delta c \Big(n,\alpha, \frac{k_c}{k}=\gamma \frac{k_{NL}}{k} , \lambda \Big)\Bigg] \nonumber
       %- \bar{c} \Big(n,\alpha, \frac{k_c}{k}=\gamma \frac{k_{NL}}{k} , \lambda \Big)\Bigg] \nonumber
\end{eqnarray}
where
\begin{equation}
\Delta c %\Big(n,\alpha, \frac{k_c}{k}=\gamma \frac{k_{NL}}{k} , \lambda \Big)=
= c \Big(n,\alpha,\gamma \frac{k_{NL}}{k} \Big)-
c \Big(n,\alpha, \lambda \Big)\,.
\end{equation}
Assuming that $\gamma \frac{k_{NL}}{k}$ and $\lambda$
are large, we can use the results of our analysis of the ultraviolet divergences in Sec. \ref{Convergence analysis} to obtain the expansion of $\Delta c$: 
\begin{eqnarray}
\label{c-regulated-full}
%\tilde{c}_{reg} (n,\alpha, k/k_{NL}) &=&
%c_{reg} (n,\alpha) 
\Delta c &=& 
%f_{-1} (\alpha) \frac{\gamma^{n+1}}{n+1} \big(\frac{k_{NL}}{k}\big)^{n+1} \nonumber \\
\frac{f_{-1}}{n+1} (\alpha) \big[\gamma^{n+1} \big(\frac{k_{NL}}{k}\big)^{n+1} -\lambda^{n+1}\big] \nonumber \\
&+&\frac{f_{1/2}}{2n-1} (\alpha) \big[\gamma^{2n-1} \big(\frac{k_{NL}}{k}\big)^{2n-1} -\lambda^{2n-1}\big]\nonumber \\
&+&\frac{f_{1}}{n-1} (\alpha) \big[\gamma^{n-1} \big(\frac{k_{NL}}{k}\big)^{n-1} -\lambda^{n-1}\big]+ \cdots \nonumber \\
\end{eqnarray}
for any $n$ other than the specific values $n=-1, 1/2, 1...$,  where the power-law
functions are replaced by logarithms. For the sake of brevity, we will not give 
results for these special cases explicitly here.  
Using these expressions the regularized one-loop result can now 
be written as
%eq.~(\ref{self-similar-dimensional-reg})
%we have 
\begin{eqnarray}
\label{sel-similar-cutoff}
\Delta_{1-loop,reg}^{2}(k) &=& \Delta_{L}^{2}(k) \Big[ 1 + c_{reg} (n,\alpha) \Delta_{L}^{2}(k) \nonumber \\
&+& f_{-1} (\alpha) \frac{\gamma^{n+1}}{n+1} (\Delta_{L}^{2}(k))^{\frac{2}{3+n}} \nonumber \\
&+& f_{1/2} (\alpha) \frac{\gamma^{2n-1}}{2n-1} (\Delta_{L}^{2}(k))^{\frac{4-n}{3+n}} \nonumber \\
&+& f_{1} (\alpha) \frac{\gamma^{n-1}}{n-1} (\Delta_{L}^{2}(k))^{\frac{4}{3+n}} \Big]
\end{eqnarray}
where 
\begin{eqnarray}
\label{c-regulated}
c_{reg} &=& \lim_{\lambda\rightarrow \infty} \Big[ c(n, \alpha, \lambda) 
- f_{-1} (\alpha) \frac{\lambda^{n+1}}{n+1} \nonumber \\
&-&  f_{1/2} (\alpha) \frac{\lambda^{2n-1}}{2n-1}   \nonumber\\
&-& f_{1} (\alpha) \frac{\lambda^{n-1}}{n-1} + \cdots \Big ]
\end{eqnarray}
This result is almost exactly equivalent to that obtained in EFT, corresponding to
the addition of the counterterms
\begin{equation}
\label{EFTcorrections}
 c_1 k^2 P_L(k) + c_2 k^4 + c_3 k^4 P_L (k)  
\end{equation}
where we have, additionally, that
\begin{eqnarray}
 c_1&=& (2\pi^2)^{\frac{2}{3+n}}f_{-1} (\alpha) \frac{\gamma^{n+1}}{n+1} \nonumber \\ 
 c_2&=&(2\pi^2)^{\frac{4-n}{n+3}}f_{1/2} (\alpha) \frac{\gamma^{2n-1}}{2n-1} \nonumber \\ 
 c_3&=&(2\pi^2)^{\frac{4}{n+3}}f_{1} (\alpha) \frac{\gamma^{n-1}}{n-1}
\end{eqnarray}
where $\gamma=k_{NL}/k_c$, a positive constant which may also 
depend also on $n$ and $\alpha$. 
{We note that 
these coefficients are predicted to be related as they are because
we have used a ``UV inspired'' strategy like that of \cite{baldauf2015bispectrum, steele2021precise}. If we used instead a symmetry-based approach, the coefficients would not be related as given, but would instead be free parameters.
}

Usually only the first two terms
in Eq.~(\ref{EFTcorrections}) are included, as they represent the leading corrections 
(the first term for $n<2$ and the second for $n>2$). As
we have discussed, this is sufficient here also other than
when $\alpha=\alpha_c$. In this case we have $c_1=0$, 
which makes the third term the leading EFT correction 
for $n<0$.
Correspondingly the expression for $c_{reg}$ is just
the unregularized result $c_{\infty} (n,\alpha)$ for 
$n<-1$, and then regularized appropriately for 
$n \geq -1$, except again for $\alpha_c$ where
the unregularized result remains valid up to
$n=1/2$. 

The ultraviolet regularized one loop result for the family of generalized scale-free models thus gives a very specific prediction that can be used in principle to test this regularization framework: the sign of the correction to the raw (unregularized) one loop result should depend on $\alpha$ as given by $f_{-1}(\alpha)$, and in particular at $\alpha=\alpha_c$ 
it vanishes so that, in this case, the raw (unregularized) one-loop result gives a well-defined finite prediction up to $n=1/2$.  
A suite of simulations for $n$ around $-1$ like those reported above for the case $n=-2$, but extending to 
smaller $\alpha$, would allow us to probe this regime. Two or higher loop corrections could also potentially 
be probed. At two loops SPT corrections diverge 
(in the ``double hard'' limit, see \cite{Baldauf2015EFTTwoLoop}) 
for $n>-2$. The coefficients of these divergences will generically be $\alpha$-dependent,  but 
we do not expect that their coefficients
will vanish at $\alpha=\alpha_c$. The regularization of these divergences in EFT will lead again to additional
terms with predicted functional dependences on $\Delta_L^2$.

Simulating small values of $\alpha$ is
however more challenging numerically. This is true because we have $k_{NL} \propto a^\frac{2 \alpha}{3+n}$, and 
therefore the ratio of the final to the initial scale factors of a simulation is given by  
\begin{equation}
\frac{a_f}{a_i}=\Big(\frac{k_{NL}(a_i)}{k_{NL}(a_f)}\Big)^\frac{3+n}{2\alpha}.
\end{equation}
In order to make use of self-similarity in order to establish accurately converged
values of the PS, we need the factor $\frac{k_{NL}(a_i)}{k_{NL}(a_f)}$ to be reasonably 
large (at least a decade). For $n=-1$ and  $\alpha\approx 0.16$ the exponent is
three times larger than it was for the smallest value $\alpha$ we reported above
for $n=-2$. This means that, for a given $\frac{k_{NL}(a_i)}{k_{NL}(a_f)}$, the 
nonlinear structures formed will become relatively much denser as $n$ increases 
and/or $\alpha$ decreases. This can in principle be remedied by using a 
sufficiently large gravitational smoothing $\epsilon$, but in this case one
must control carefully that its effects do not propagate to the intermediate
(weakly nonlinear) scales we are interested in for comparison with perturbation
theory.

\section{Discussion and conclusions}

We have studied the PS calculated at one loop in standard Eulerian perturbation theory for the family of generalized scale-free cosmologies, characterized by initial Gaussian fluctuations with a pure power law PS and an EdS expansion driven by clustering pressureless matter as well as a  smooth pressureless component. We have thus generalized existing analytic results for the standard EdS case 
\cite{makino1992,scoccimarro1996loop} to this one-parameter family, with the corresponding analytic expressions now becoming functions not just of the power law exponent $n$ but also of the logarithmic growth rate $\alpha$ in the model. While in the standard case $\alpha=1$, the parameter $\alpha$ can vary in the range $0< \alpha \leq \infty$, with the lower limit corresponding to an infinite Hubble rate and the upper limit to a static universe. 

While these models are idealized and very different from typical standard cosmological models, they provide a
simple framework in which to test cosmological perturbation theory. Specifically they are evidently 
designed to probe the cosmology dependence, and indeed we have seen here that, by exploiting self-similarity, 
it is possible with even quite small $N$-body simulations to test and validate  its predictions to a high degree of accuracy. To our knowledge, this is the first time that the predictions of perturbation theory for  
dependence on the growth rate of fluctuations have been tested numerically.

Further we have argued that these models are an interesting tool to probe the 
regularization of PT, and specifically the EFT approach to this problem. This is the case
because the associated divergences, and thus their regularization, have nontrivial 
dependences on the parameter $\alpha$. In particular this leads to the vanishing 
of the leading correction in EFT at a specific value of $\alpha$.  
As we have discussed the regime of $n$ and small $\alpha$ 
of relevance to test these predictions poses some numerical challenges beyond that 
which was needed for the simulations we have reported here. This is the subject of
ongoing study.

\begin{acknowledgments}
We thank 
Bruno Marcos for his collaboration in the modification of GADGET2 in \cite{benhaiem2014self}, and Sara Maleubre for useful discussions
on the analysis of the PS in scale-free simulations.
A.P. is supported by Indonesia Endowment Fund for Education (LPDP).
Numerical simulations have
been performed on a cluster at MeSU hosted at
Sorbonne Université.
\end{acknowledgments}

\bibliographystyle{apsrev4-2}
\bibliography{references}% Produces the bibliography via BibTeX.
\end{document}